\documentclass[twocolumn,superscriptaddress,amsmath,amssymb,floats]{revtex4-1}
\usepackage[latin9]{inputenc}
\usepackage{color}
\usepackage{mathtools}
\usepackage{amsmath}
\usepackage{amssymb}
\usepackage{graphicx}
\usepackage{dsfont}

\makeatletter

\newcommand*\LyXThinSpace{\,\hspace{0pt}}

\@ifundefined{textcolor}{}
{%
 \definecolor{BLACK}{gray}{0}
 \definecolor{WHITE}{gray}{1}
 \definecolor{RED}{rgb}{1,0,0}
 \definecolor{GREEN}{rgb}{0,1,0}
 \definecolor{BLUE}{rgb}{0,0,1}
 \definecolor{CYAN}{cmyk}{1,0,0,0}
 \definecolor{MAGENTA}{cmyk}{0,1,0,0}
 \definecolor{YELLOW}{cmyk}{0,0,1,0}
}

\usepackage{epstopdf}
\usepackage{array}
\usepackage{mathrsfs}
\usepackage{dcolumn}
\usepackage{bm}

\newcolumntype{L}[1]{>{\raggedright\let\newline\\\arraybackslash\hspace{0pt}}m{#1}}
\newcolumntype{C}[1]{>{\centering\let\newline\\\arraybackslash\hspace{0pt}}m{#1}}
\newcolumntype{R}[1]{>{\raggedleft\let\newline\\\arraybackslash\hspace{0pt}}m{#1}}

\newcommand{\mbf}[1]{\mathbf{#1}}

\usepackage{tikz}
\usetikzlibrary{calc,intersections}
\usetikzlibrary{shadings}
\usepgflibrary{shadings}

\makeatother

\begin{document}

\title{Spin reorientation driven by the interplay between spin-orbit coupling
and Hund's rule coupling in iron pnictides}

\author{Morten H. Christensen}

\affiliation{School of Physics and Astronomy, University of Minnesota, Minneapolis,
MN 55455, USA}

\affiliation{Niels Bohr Institute, University of Copenhagen, DK-2100, Denmark}

\author{Jian Kang}

\affiliation{School of Physics and Astronomy, University of Minnesota, Minneapolis,
MN 55455, USA}

\author{Brian M. Andersen}

\affiliation{Niels Bohr Institute, University of Copenhagen, DK-2100, Denmark}

\author{Ilya Eremin}

\affiliation{Institut f{\"u}r Theoretische Physik III, Ruhr-Universit{\"a}t Bochum,
44801 Bochum, Germany}

\affiliation{National University of Science and Technology `MISiS', 119049 Moscow, Russian Federation}

\author{Rafael M. Fernandes}

\affiliation{School of Physics and Astronomy, University of Minnesota, Minneapolis,
MN 55455, USA}
\begin{abstract}
In most magnetically-ordered iron pnictides, the magnetic moments
lie in the FeAs planes, parallel to the modulation direction of the
spin stripes. However, recent experiments in hole-doped iron pnictides
have observed a reorientation of the magnetic moments from in-plane
to out-of-plane. Interestingly, this reorientation is accompanied
by a change in the magnetic ground state from a stripe antiferromagnet
to a tetragonal non-uniform magnetic configuration. Motivated by these
recent observations, here we investigate the origin of the spin anisotropy
in iron pnictides using an itinerant microscopic electronic model
that respects all the symmetry properties of a single FeAs plane.
We find that the interplay between the spin-orbit coupling and the
Hund's rule coupling can account for the observed spin anisotropies,
including the spin reorientation in hole-doped pnictides, without
the need to invoke orbital or nematic order. Our calculations also
reveal an asymmetry between the magnetic ground states of electron-
and hole-doped compounds, with only the latter displaying tetragonal
magnetic states. 
\end{abstract}
\maketitle

\section{Introduction}

In the iron pnictides, unconventional superconductivity appears in
close proximity to a magnetic instability \cite{hosono02,greene01,chubukov01,hirschfeld01}.
As a result, much of the research into these compounds has been devoted
to understanding the magnetic properties of these systems \cite{dagotto01,mazin01,dai01,inosov02}.
Experimentally, the spin-density wave (SDW) magnetic order of most
iron pnictides has orthorhombic ($C_{2}$) symmetry and corresponds
to stripes of parallel spins modulated either along the $\hat{\mathbf{x}}$
direction (i.e. ordering vector $\mathbf{Q}_{1}=\left(\pi,0\right)$
and staggered magnetic order parameter $\mathbf{M}_{1}$) or along
the $\hat{\mathbf{y}}$ direction (i.e. ordering vector $\mathbf{Q}_{2}=\left(0,\pi\right)$
and staggered magnetic order parameter $\mathbf{M}_{2}$), in the
coordinate system of the Fe square lattice \cite{inosov02,dai01}.
Theoretically, this state has been described by a variety of approaches,
from purely localized Heisenberg spins \cite{kivelson01,kruger01,sachdev01,abrahams01,batista01}
to itinerant nesting-based scenarios \cite{fernandes01,eremin01,cvetkovic03,brydon01,lorenzana01,kovacic01,chubukov08,klauss08,knolle10}
to hybrid models mixing local moments and itinerant carriers \cite{dagotto01,ku02,phillips01,fernandes03,johannes01,liang13}.
Common to nearly all these approaches is the assumption that the magnetic
degrees of freedom have an underlying $O(3)$ spin-rotational symmetry.
From a phenomenological perspective, this implies that the magnetic
free energy $F_{\mathrm{mag}}$ depends only on the absolute value
of the magnetic order parameters, i.e. $F_{\mathrm{mag}}\left(M_{1}^{2},M_{2}^{2}\right)$
\cite{fanfarillo01}.

Despite the success of these approaches in describing many magnetic
properties of the iron pnictides -- such as the onset of a preemptive
nematic transition \cite{fernandes02} and the appearance of a tetragonal
magnetic ground state \cite{lorenzana01} -- there are important features
that remain largely unaddressed. In particular, the $O(3)$ rotational
symmetry of a free spin does not hold for a magnetic moment in a crystal.
Instead, the symmetries of the underlying lattice induce anisotropies
in spin space that may be significant \cite{devereaux01}. Indeed,
in most iron pnictides, the magnetic moments are observed to point
parallel to the modulation vector of the stripes, i.e. $\mathbf{M}_{i}\parallel\mathbf{Q}_{i}$
\cite{dai01,inosov02}. Attesting the significance of this spin anisotropy,
a sizable spin gap of the order of $10$ meV is also found at low
temperatures deep in the magnetically ordered state \cite{Park12,tucker01,dai02,dai03}.
Interestingly, recent experiments in hole-doped iron pnictides have
reported a spin reorientation near optimal doping, in which the direction
of the magnetic moments flip from in-plane to out-of-plane \cite{waser01,osborn01,allred01,allred02}.
Remarkably, this spin reorientation takes place in a region of the
phase diagram in which the magnetic ground state changes from stripe/orthorhombic
to tetragonal.

Therefore, elucidating the origins of these spin anisotropies and
their impact on the normal state properties is paramount to advance
our understanding of the iron pnictides. A natural candidate to account
for these effects is the spin-orbit coupling (SOC) term $\lambda\mbf{S}\cdot\mbf{L}$
\cite{cvetkovic01,fernandes04,hirschfeld02,si01}, which converts
the lattice anisotropies into anisotropies in spin space. Recent ARPES
measurements of the SOC $\lambda$ have reported values of the order
of $20$ meV \cite{borisenko01}, which is not far from the typical
magnetic energy scale of the problem (as extracted for instance from
optical conductivity measurements \cite{nakajima01,degiorgi01}).
To include the SOC term in theoretical models, it is necessary to
account for the puckering of the As atoms along the FeAs plane, which
effectively doubles the unit cell of the Fe-only square lattice. In
this paper, instead of working with the cumbersome ten-band model
relevant for the 2-Fe unit cell, we consider a simpler low-energy
microscopic model that respects all the symmetries of the FeAs plane
and focuses only on the states near the Fermi level. Such a model,
which relies on the smallness of the Fermi surface pockets of the
iron pnictides, was previously derived by Cvetkovic and Vafek using
rigorous group theoretical arguments \cite{cvetkovic01}. Here, we
show how the main ingredients of the model can be derived from a straightforward
expansion of the usual five-orbital model for the pnictides. By computing
microscopically the magnetic free energy in the paramagnetic state,
we find the leading-order magnetic anisotropic terms: 
\begin{eqnarray}
\delta F & = & \alpha_{1}\left(M_{1,x}^{2}+M_{2,y}^{2}\right)\nonumber \\
 &  & +\alpha_{2}\left(M_{1,y}^{2}+M_{2,x}^{2}\right)\nonumber \\
 &  & +\alpha_{3}\left(M_{1,z}^{2}+M_{2,z}^{2}\right)\,.
\end{eqnarray}

The anisotropic coefficients $\alpha_{i}$ are proportional not only
to the square of the SOC term, $\lambda^{2}$, but also to the Hund's
rule coupling $J$. Evaluation of the coefficients reveals that $\alpha_{1}<\alpha_{3},\alpha_{2}$
for most of the temperature-doping phase diagram, implying that the
magnetic moments have a general tendency to lie in the plane. Interestingly,
in the hole-doped side of the phase diagram, we find a small region
in which $\alpha_{3}<\alpha_{1},\alpha_{2}$, indicating a spin reorientation
from in-plane to out-of-plane. Both results are in qualitative agreement
with the observations discussed above, providing evidence that the
SOC term, with the aid of the Hund's rule coupling, is sufficient
to account for the magnetic anisotropies of the iron pnictides. This
conclusion contrasts with previous proposals that orbital and/or nematic
order are necessary to explain the observed magnetic moment orientation
\cite{devereaux01}.

For completeness, we also analyze the nature of the magnetic ground
state across the phase diagram. We find a general tendency of electron-doped
compounds to form an orthorhombic uniaxial (single-$\mbf{Q}$) stripe
state (i.e. either $\left\langle \left|\mathbf{M}_{1}\right|\right\rangle =0$
or $\left\langle \left|\mathbf{M}_{2}\right|\right\rangle =0$), whereas
hole-doped compounds favor a tetragonal biaxial (double-$\mbf{Q}$)
magnetic state (i.e. $\left\langle \left|\mathbf{M}_{1}\right|\right\rangle =\left\langle \left|\mathbf{M}_{2}\right|\right\rangle $).
Such an electron-hole asymmetry is also qualitatively consistent with
experiments -- and in particular with the recent observation that
the spin reorientation takes place in a region of the phase diagram
in which the magnetic ground state is tetragonal.

The paper is organized as follows: In section \ref{sec:model} we
introduce the low-energy microscopic model with the SOC term and the
electronic interactions. Section \ref{sec:second_order_free_energy}
is devoted to the analysis of the coefficients of the free energy
responsible for the magnetic anisotropy within leading-order. In section
\ref{sec:c2vsc4} we refine the phase diagram by including fourth
order contributions to the free energy that allow us to distinguish
between stripe and tetragonal magnetic ground states. Concluding remarks
are presented in section \ref{sec:conclusions}. Details of the calculations
are included in four appendices.

\section{Low-energy microscopic model}

\label{sec:model}

\begin{figure}
\centering \includegraphics[width=0.45\textwidth]{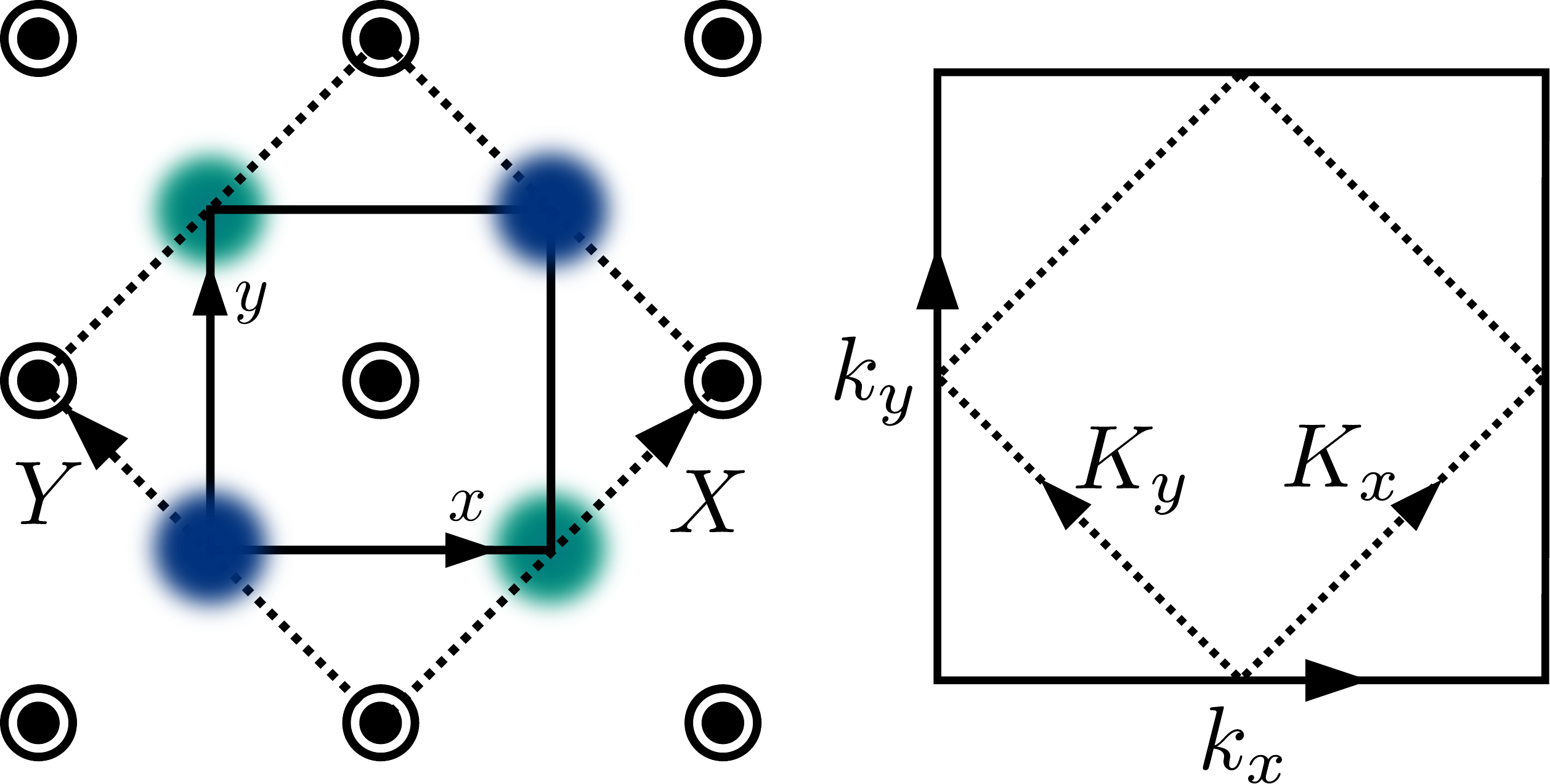}
\protect\protect\protect\caption{\label{fig:unit_cells} (Color online) (Left) Illustration of the
2-Fe (dotted line) and 1-Fe (solid line) unit cells. The black dots
denote iron atoms, while the pnictogens form two sublattices, one
above the iron-plane (dark blue) and one below the iron-plane (light
blue). (Right) Brillouin zones corresponding to the 1-Fe and 2-Fe
unit cells. The dotted line is the ``folded'' Brillouin zone, corresponding
to the 2-Fe unit cell, while the solid line is the ``unfolded''
Brillouin zone, corresponding to the 1-Fe unit cell.}
\end{figure}

We start with a low-energy microscopic model that focuses only on
the electronic states near the Fermi level, while respecting the symmetries
of the FeAs plane. Such a model was originally derived in Ref. \cite{cvetkovic01}
using the symmetry properties of the non-symmorphic space group $P4/nmm$
of a single FeAs plane. Here, we present an alternative derivation
based on the typical 5-orbital tight-binding model used for the iron
pnictides \cite{graser01}: 
\begin{equation}
\mathcal{H}_{0}=\sum_{\mathbf{k}\mu\nu\alpha}\varepsilon_{\mu\nu}\left(\mathbf{k}\right)c_{\mu,\mathbf{k}\alpha}^{\dagger}c_{\nu,\mathbf{k}\alpha}^{\phantom{\dagger}}\label{H0}
\end{equation}
where $\mathbf{k}$ is the momentum, $\alpha$ is the spin, and $\mu,\nu$
denote one of the five Fe orbitals, $xz$, $yz$, $x^{2}-y^{2}$,
$xy$, and $3z^{2}-r^{2}$. The matrix $\varepsilon_{\mu\nu}\left(\mathbf{k}\right)$
corresponds to the Fourier-transformed tight-binding dispersions involving
up to fourth-nearest neighbor hoppings. Its explicit expression is
given in Appendix A. Note that this Hamiltonian is based on the single-Fe
square lattice (i.e. it refers to the ``unfolded'' Brillouin zone),
and that the coordinate system is defined such that $k_{x}$ and $k_{y}$
are parallel to the nearest-neighbor Fe atoms directions. The actual
crystallographic unit cell contains two Fe atoms due to the puckering
of the As atoms, resulting in the so-called ``folded'' Brillouin
zone, described by the coordinates $K_{x},K_{y}$ (see Fig. \ref{fig:unit_cells}).
Note that the two coordinate systems are related by: 
\begin{align}
K_{x} & =k_{x}+k_{y}\nonumber \\
K_{y} & =-k_{x}+k_{y}\label{coordinate_change}
\end{align}
where the momentum in the unfolded zone is measured in units of its
inverse lattice constant $1/a$, whereas the momentum in the folded
zone is measured in units of the its inverse lattice constant $1/\left(\sqrt{2}a\right)$.

The key properties that allow us to derive a simpler low-energy model
are the facts that the Fermi surface pockets are small and that the
orbitals that mostly contribute to the Fermi surface are $xz$, $yz$,
$xy$. In particular, the idea is to start at the high-symmetry points
of the unfolded Brillouin zone (namely, $\Gamma=\left(0,0\right)$,
$X=\left(\pi,0\right)$, and $Y=\left(0,\pi\right)$), where the band
states are pure orbital states, and perform an expansion of the corresponding
matrix elements $\varepsilon_{\mu\nu}\left(\mathbf{k}\right)$ for
small momentum. Note that, to focus on a general and analytically
tractable model, we follow Ref. \cite{cvetkovic01} and ignore the
states near the $(\pi,\pi)$ point of the unfolded Brillouin zone.
While it is true that some iron pnictides display a hole-pocket with
$xy$-orbital character centered at this point, this pocket is not
usually present for all values of $k_{z}$, and is absent in many
of the iron-based materials with a single FeAs plane per unit cell.
Correspondingly we consider in this work the doping range in which
this pocket lies below the Fermi level. 

Consider first the $\Gamma$ point; the two states closest to the
Fermi level are the $xz$ and $yz$ orbitals, which form a degenerate
doublet in the absence of SOC. Thus, for small $\mathbf{k}$, we define
the spinor: 
\begin{equation}
\psi_{\Gamma,\mathbf{k}}=\left(\begin{array}{c}
\phantom{-}c_{yz,\mathbf{k}\uparrow}\\
-c_{xz,\mathbf{k}\uparrow}\\
\phantom{-}c_{yz,\mathbf{k}\downarrow}\\
-c_{xz,\mathbf{k}\downarrow}
\end{array}\right)\,.\label{psi_Gamma}
\end{equation}

Projecting $\varepsilon_{\mu\nu}\left(\mathbf{k}\right)$ on this
sub-space and expanding for small $\mathbf{k}$ then yields the $4\times4$
Hamiltonian: 
\begin{equation}
H_{0,\Gamma}=\sum_{\mathbf{k}}\psi_{\Gamma,\mathbf{k}}^{\dagger}h_{\Gamma}\left(\mathbf{k}\right)\psi_{\Gamma,\mathbf{k}}^{\phantom{\dagger}}\label{H0_Gamma}
\end{equation}
with 
\begin{align}
 & h_{\Gamma}(\mbf{k})=\nonumber \\
  & \begin{psmallmatrix}\epsilon_{\Gamma}+2\frac{\mbf{k}^{2}}{2m_{\Gamma}}+b\left(k_{x}^{2}-k_{y}^{2}\right) & 4ck_{x}k_{y}\\
4ck_{x}k_{y} & \epsilon_{\Gamma}+2\frac{\mbf{k}^{2}}{2m_{\Gamma}}-b\left(k_{x}^{2}-k_{y}^{2}\right)
\end{psmallmatrix}\otimes\sigma^{0}\label{aux_H0_Gamma}
\end{align}
where $\sigma^{0}$ is a Pauli matrix acting on spin space. The coefficients
$\epsilon_{\Gamma}$, $m_{\Gamma}$, $b$, and $c$ can be obtained
directly from the tight-binding parameters (see Appendix A). Note,
however, that as we move away from the high-symmetry points of the
Brillouin zone, other orbitals start to contribute to the electronic
states. Consequently, the coefficients of the expansion (as derived
in Appendix \ref{app:heurestic_model}) will be slightly renormalized
by the hybridization with the orbitals not included in the expansion,
although the form of the expansion remains invariant. To account for
this issue, we can consider the coefficients to be free parameters
that can be fit directly to the first-principle band dispersions.

Near the $X$ point, the low-energy states correspond to the orbitals
$yz$ and $xy$. Defining the spinor: 
\begin{equation}
\psi_{X,\mathbf{k}+\mathbf{Q}_{1}}=\left(\begin{array}{c}
c_{yz,\mathbf{k}+\mathbf{Q}_{1}\uparrow}\\
c_{xy,\mathbf{k}+\mathbf{Q}_{1}\uparrow}\\
c_{yz,\mathbf{k}+\mathbf{Q}_{1}\downarrow}\\
c_{xy,\mathbf{k}+\mathbf{Q}_{1}\downarrow}
\end{array}\right)\,.\label{psi_X}
\end{equation}
and expanding the projected $\varepsilon_{\mu\nu}\left(\mathbf{k}\right)$
near $\mathbf{Q}_{1}=\left(\pi,0\right)$ yields: 
\begin{equation}
H_{0,X}=\sum_{\mathbf{k}}\psi_{X,\mathbf{k}+\mathbf{Q}_{1}}^{\dagger}h_{X}\left(\mathbf{k}+\mathbf{Q}_{1}\right)\psi_{X,\mathbf{k}+\mathbf{Q}_{1}}^{\phantom{\dagger}}\label{H0_X}
\end{equation}
with 
\begin{align}
 & h_{X}(\mathbf{k}+\mathbf{Q}_{1})=\nonumber \\
 & \begin{psmallmatrix}\epsilon_{1}+2\frac{\mbf{k}^{2}}{2m_{1}}+a_{1}\left(k_{x}^{2}-k_{y}^{2}\right) & -iv_{X}(\mbf{k})\\
iv_{X}(\mbf{k}) & \epsilon_{3}+2\frac{\mbf{k}^{2}}{2m_{3}}+a_{3}\left(k_{x}^{2}-k_{y}^{2}\right)
\end{psmallmatrix}\otimes\sigma^{0}\label{aux_H0_X}
\end{align}
and: 
\begin{equation}
v_{X}(\mbf{k})=2vk_{y}+2p_{1}k_{y}(k_{y}^{2}+3k_{x}^{2})-2p_{2}k_{y}(k_{x}^{2}-k_{y}^{2})\,.\label{vX}
\end{equation}

Similarly, near the $Y$ point, the low-energy states involve the
orbitals $xz$ and $xy$: 
\begin{equation}
\psi_{Y,\mathbf{k}+\mathbf{Q}_{2}}=\left(\begin{array}{c}
c_{xz,\mathbf{k}+\mathbf{Q}_{2}\uparrow}\\
c_{xy,\mathbf{k}+\mathbf{Q}_{2}\uparrow}\\
c_{xz,\mathbf{k}+\mathbf{Q}_{2}\downarrow}\\
c_{xy,\mathbf{k}+\mathbf{Q}_{2}\downarrow}
\end{array}\right)\,.\label{psi_Y}
\end{equation}
Projecting and expanding $\varepsilon_{\mu\nu}\left(\mathbf{k}\right)$
near $\mathbf{Q}_{2}=\left(0,\pi\right)$ gives: 
\begin{equation}
H_{0,Y}=\sum_{\mathbf{k}}\psi_{Y,\mathbf{k}+\mathbf{Q}_{2}}^{\dagger}h_{Y}\left(\mathbf{k}+\mathbf{Q}_{2}\right)\psi_{Y,\mathbf{k}+\mathbf{Q}_{2}}^{\phantom{\dagger}}\label{H0_Y}
\end{equation}
with 
\begin{align}
 & h_{Y}(\mathbf{k}+\mathbf{Q}_{2})=\nonumber \\
 & \begin{psmallmatrix}\epsilon_{1}+2\frac{\mbf{k}^{2}}{2m_{1}}-a_{1}\left(k_{x}^{2}-k_{y}^{2}\right) & -iv_{Y}(\mbf{k})\\
iv_{Y}(\mbf{k}) & \epsilon_{3}+2\frac{\mbf{k}^{2}}{2m_{3}}-a_{3}\left(k_{x}^{2}-k_{y}^{2}\right)
\end{psmallmatrix}\otimes\sigma^{0}\label{aux_H0_Y}
\end{align}
and: 
\begin{equation}
v_{Y}(\mbf{k})=-2vk_{x}-2p_{1}k_{x}(k_{x}^{2}+3k_{y}^{2})-2p_{2}k_{x}(k_{x}^{2}-k_{y}^{2})\,.\label{vY}
\end{equation}

Having established the low-energy states in the unfolded Brillouin
zone (i.e. the one referring to the 1-Fe unit cell), it is now straightforward
to fold the states into the 2-Fe unit cell (see Fig. \ref{fig:unit_cells}).
Despite working in the folded Brillouin zone, described by the coordinates
$K_{x},K_{y}$, we will still make use of the coordinates $k_{x},k_{y}$
of the unfolded zone. From Eq. (\ref{coordinate_change}), we find
that upon folding, both momenta $\mathbf{Q}_{1}=\left(\pi,0\right)$
and $\mathbf{Q}_{2}=\left(0,\pi\right)$ are identified with the same
momentum $\mathbf{Q}_{M}=\left(\pi,\pi\right)$. It is straightforward
to show that the spinors $X$ and $Y$ now combine to form two new
degenerate doublets at the $M=\left(\pi,\pi\right)$ point of the
folded zone: 
\begin{equation}
\psi_{M_{1},\mathbf{k}+\mathbf{Q}_{M}}=\left(\begin{array}{c}
c_{xz,\mathbf{k}+\mathbf{Q}_{2}\uparrow}\\
c_{yz,\mathbf{k}+\mathbf{Q}_{1}\uparrow}\\
c_{xz,\mathbf{k}+\mathbf{Q}_{2}\downarrow}\\
c_{yz,\mathbf{k}+\mathbf{Q}_{1}\downarrow}
\end{array}\right)\,;\:\psi_{M_{3},\mathbf{k}+\mathbf{Q}_{M}}=\left(\begin{array}{c}
c_{xy,\mathbf{k}+\mathbf{Q}_{2}\uparrow}\\
c_{xy,\mathbf{k}+\mathbf{Q}_{1}\uparrow}\\
c_{xy,\mathbf{k}+\mathbf{Q}_{2}\downarrow}\\
c_{xy,\mathbf{k}+\mathbf{Q}_{1}\downarrow}
\end{array}\right)\label{psi_M}
\end{equation}

\begin{figure}[t]
\centering \includegraphics[width=0.45\textwidth]{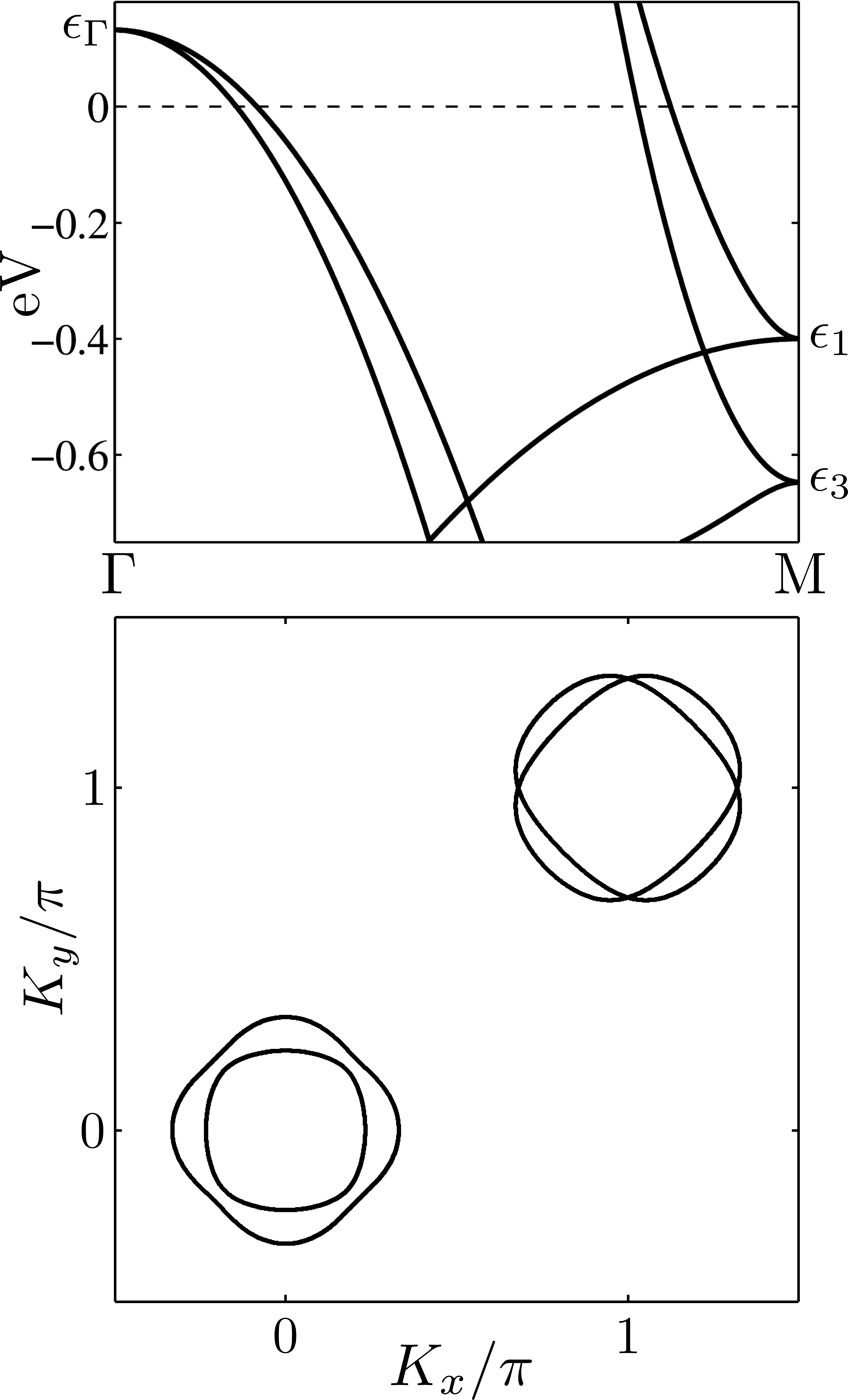}
\protect\protect\protect\protect\caption{\label{fig:band_structure}(upper panel) Cut of the low-energy band
dispersion from the $\Gamma=\left(0,0\right)$ to the $M=\left(\pi,\pi\right)$
point of the folded Brillouin zone with parameters fit to the tight-binding
model of Ref. \cite{cvetkovic02}. The corresponding Fermi surface
is shown in the lower panel.}
\end{figure}

Hereafter, we will consider the momentum of any spinor as measured
relative to the high-symmetry points, as appropriate. Then, the non-interacting
Hamiltonian becomes: 
\begin{equation}
\mathcal{H}_{0}=\sum_{\mbf{k}}\Psi_{\mathbf{k}}^{\dagger}\left[H_{0}(\mbf{k})-\mu\mathds{1}\right]\Psi_{\mathbf{k}}\,,\label{H0_final_qux}
\end{equation}
where we defined the enlarged spinor: 
\begin{equation}
\Psi_{\mathbf{k}}=\left(\begin{array}{c}
\psi_{Y,\mathbf{k}}\\
\psi_{X,\mathbf{k}}\\
\psi_{\Gamma,\mathbf{k}}
\end{array}\right)\label{aux_spinor}
\end{equation}
and the Hamiltonian matrix: 
\begin{equation}
H_{0}(\mbf{k})=\begin{pmatrix}h_{Y}(\mbf{k}) & 0 & 0\\
0 & h_{X}(\mbf{k}) & 0\\
0 & 0 & h_{\Gamma}(\mbf{k})
\end{pmatrix}\label{aux_H0_final}
\end{equation}
where $\mu$ is the chemical potential and $\mathds{1}$ is the identity matrix.
We note that this model has the same properties of the Hamiltonian
derived by Cvetkovic and Vafek in Ref. \cite{cvetkovic01} combining
a $\mathbf{k}\cdot\mathbf{p}$ expansion and the symmetry properties
of the $P4/nmm$ space group (note, however, that the definition of
the spinors $X$ and $Y$ are switched in Ref. \cite{cvetkovic01}
with respect to the notation adopted here). In the group-theory language,
the spinor $\psi_{\Gamma}$ belongs to the two-dimensional $E_{g}$
representation of $P4/nmm$ near the $\Gamma$ point, whereas $\psi_{M_{1}}$
and $\psi_{M_{3}}$ belong to the two-dimensional $E_{M_{1}}$ and
$E_{M_{3}}$ representations of $P4/nmm$ near the $M$ point. Hereafter,
we will use for the coefficients of the Hamiltonian the parameters
given by Table IX in Ref. \cite{cvetkovic01}. Those were obtained
by direct fitting of the band dispersions to first-principle calculations.
The resulting band dispersions, as well as the Fermi surface, are
shown in Fig. \ref{fig:band_structure}. Note also
that this low-energy model is fundamentally different than two-orbital
models that restrict the Hamiltonian to the subspace of the $xz$
and $yz$ orbitals. Our model, derived from the five-orbital tight-binding
model as shown in Appendix A, not only obeys all the symmetries imposed
by the $P4/nmm$ space group, but it also contains information about
all the orbitals that contribute to the Fermi surface, including the
$xy$ orbital.

Besides the band dispersions, the non-interacting Hamiltonian must
also contain the SOC term $\lambda\mathbf{S}\cdot\mathbf{L}$, with
$\mbf{S}$ denoting the spin angular momentum operator and $\mbf{L}$,
the orbital angular momentum operator. Note that this atomic-like
term preserves the Kramers degeneracy of each state. To proceed, we
project this term from the $L=2$ cubic harmonic basis to the orbital
basis (see Appendix B for more details). At the $\Gamma$ point, we
obtain an admixture of the $xz$ and $yz$ orbitals: 
\begin{align}
\frac{\lambda}{2}\sum_{\mathbf{k}\alpha\beta}\left(ic_{yz,\mathbf{k}\alpha}^{\dagger}\sigma_{\alpha\beta}^{z}c_{xz,\mathbf{k}\beta}^{\phantom{\dagger}}+\mathrm{h.c.}\right) & =\nonumber \\
\frac{\lambda}{2}\sum_{\mathbf{k}}\psi_{\Gamma,\mathbf{k}}^{\dagger}\left(\tau^{y}\otimes\sigma^{z}\right)\psi_{\Gamma,\mathbf{k}}^{\phantom{\dagger}}\label{Gamma_SOC}
\end{align}
where, in the last step, we used the definition of the spinors. At
the $M$ point, we obtain the admixture of $xz/yz$ and $xy$ orbitals:
\begin{align}
\frac{\lambda}{2}\sum_{\mathbf{k}\alpha\beta}\left(ic_{xz,\mathbf{k}\alpha}^{\dagger}\sigma_{\alpha\beta}^{x}c_{xy,\mathbf{k}\beta}^{\phantom{\dagger}}+\mathrm{h.c.}\right) & =\nonumber \\
\frac{\lambda}{2}\sum_{\mathbf{k}}\left[i\psi_{Y,\mathbf{k}+\mathbf{Q}_{2}}^{\dagger}\left(\tau^{+}\otimes\sigma^{x}\right)\psi_{X,\mathbf{k}+\mathbf{Q}_{1}}^{\phantom{\dagger}}+\mathrm{h.c.}\right]\label{M1_SOC}
\end{align}
as well as: 
\begin{align}
\frac{\lambda}{2}\sum_{\mathbf{k}\alpha\beta}\left(ic_{xy,\mathbf{k}\alpha}^{\dagger}\sigma_{\alpha\beta}^{y}c_{yz,\mathbf{k}\beta}^{\phantom{\dagger}}+\mathrm{h.c.}\right) & =\nonumber \\
\frac{\lambda}{2}\sum_{\mathbf{k}}\left[i\psi_{Y,\mathbf{k}+\mathbf{Q}_{2}}^{\dagger}\left(\tau^{-}\otimes\sigma^{y}\right)\psi_{X,\mathbf{k}+\mathbf{Q}_{1}}^{\phantom{\dagger}}+\mathrm{h.c.}\right]\label{M2_SOC}
\end{align}
with $\tau^{\pm}=\frac{1}{2}\left(\tau^{x}\pm i\tau^{y}\right)$.
Therefore, the SOC becomes: 
\begin{equation}
\mathcal{H}_{\mathrm{SOC}}=\sum_{\mbf{k}}\Psi_{\mathbf{k}}^{\dagger}H_{\mathrm{SOC}}(\mbf{k})\Psi_{\mathbf{k}}\,,\label{H0_final}
\end{equation}
with: 
\begin{equation}
H_{\mathrm{SOC}}(\mbf{k})=\begin{pmatrix}0 & h_{M}^{\text{SOC}}(\mbf{k}) & 0\\
\left(h_{M}^{\text{SOC}}(\mbf{k})\right)^{\dagger} & 0 & 0\\
0 & 0 & h_{\Gamma}^{\text{SOC}}(\mbf{k})
\end{pmatrix}\label{H_SOC}
\end{equation}
such that: 
\begin{eqnarray}
h_{\Gamma}^{\text{SOC}}(\mbf{k}) & = & \frac{1}{2}\lambda\left(\tau^{y}\otimes\sigma^{z}\right)\,,\\
h_{M}^{\text{SOC}}(\mbf{k}) & = & \frac{i}{2}\lambda\left(\tau^{+}\otimes\sigma^{x}+\tau^{-}\otimes\sigma^{y}\right)\,,
\end{eqnarray}
in agreement with the group-theoretical arguments of Ref. \cite{cvetkovic01}.

The interacting part of this low-energy model is rather complex, involving
30 different possible biquadratic terms in the fermionic operators.
Here, we will focus on the interactions coupling the $\Gamma$ and
the $M$ points, since those are the ones that will be relevant for
the calculation of the magnetic action in the next section. Defining
$\tilde{\tau}^{1,3}\equiv\tfrac{1}{2}(\tau^{0}\pm\tau^{z})$, the
interacting terms coupling the $\Gamma$ and $M$ points are written
as (see Ref. \cite{cvetkovic01}):

\begin{eqnarray}
\mathcal{H}_{\text{int}} & = & \frac{1}{2}\sum_{\mbf{k}\sigma}\bigg[v_{13}\left(\psi_{X\sigma}^{\dagger}(\mbf{k})\tau^{-}\psi_{\Gamma\sigma}(\mbf{k})+\mathrm{h.c.}\right)^{2}\nonumber \\
 &  & +v_{13}\left(\psi_{Y\sigma}^{\dagger}(\mbf{k})\tilde{\tau}^{3}\psi_{\Gamma\sigma}(\mbf{k})+\mathrm{h.c.}\right)^{2}\nonumber \\
 &  & +v_{15}\left(\psi_{X\sigma}^{\dagger}(\mbf{k})\tilde{\tau}^{3}\psi_{\Gamma\sigma}(\mbf{k})+\mathrm{h.c.}\right)^{2}\nonumber \\
 &  & +v_{15}\left(\psi_{Y\sigma}^{\dagger}(\mbf{k})\tau^{-}\psi_{\Gamma\sigma}(\mbf{k})+\mathrm{h.c.}\right)^{2}\nonumber \\
 &  & +v_{17}\left(\psi_{X\sigma}^{\dagger}(\mbf{k})\tau^{+}\psi_{\Gamma\sigma}(\mbf{k})+\mathrm{h.c.}\right)^{2}\nonumber \\
 &  & +v_{17}\left(\psi_{Y\sigma}^{\dagger}(\mbf{k})\tilde{\tau}^{1}\psi_{\Gamma\sigma}(\mbf{k})+\mathrm{h.c.}\right)^{2}\nonumber \\
 &  & +v_{19}\left(\psi_{X\sigma}^{\dagger}(\mbf{k})\tilde{\tau}^{1}\psi_{\Gamma\sigma}(\mbf{k})+\mathrm{h.c.}\right)^{2}\nonumber \\
 &  & +v_{19}\left(\psi_{Y\sigma}^{\dagger}(\mbf{k})\tau^{+}\psi_{\Gamma\sigma}(\mbf{k})+\mathrm{h.c.}\right)^{2}\bigg]\,,\label{eq:hamilton_interaction}
\end{eqnarray}

Note that all terms are diagonal in spin space. In terms of the more
usual multi-orbital Hubbard model with onsite interactions, the first
three coefficients originate from the Hund's rule coupling, $v_{13}=v_{15}=v_{17}=J$,
while the last one arises from the intra-orbital Hubbard term, $v_{19}=U/2$
\cite{cvetkovic01}. Here, we are not interested in which interactions
will drive the SDW transition. Rather, we will assume a nearby SDW
instability and compute how the interplay between these interactions
and the SOC affect the magnetic action.

Finally, in order to be able to derive the magnetic action in the
next sections, we need also to establish how the magnetic order parameters
$\mbf{M}_{1}$ and $\mbf{M}_{2}$, corresponding to $(\pi,0)$ and
$(0,\pi)$ order in the unfolded zone, couple to the low-energy electronic
states. Here, we will consider only intra-orbital magnetism. Indeed,
previous Hartree-Fock investigations of the five-orbital Hubbard model
have shown that the dominant contributions to the magnetic instability
arise from intra-orbital couplings \cite{gastiasoro01,daghofer01}.
Therefore, the SDW vertices become: 
\begin{align}
\mathcal{H}_{\mathrm{SDW}}=\mathbf{M}_{1}\cdot\sum_{\mathbf{k}\alpha\beta}\left(c_{yz,\mathbf{k}\alpha}^{\dagger}\boldsymbol{\sigma}{}_{\alpha\beta}c_{yz,\mathbf{k}+\mathbf{Q}_{1}\beta}^{\phantom{\dagger}}+\mathrm{h.c.}\right)\nonumber \\
+\mathbf{M}_{2}\cdot\sum_{\mathbf{k}\alpha\beta}\left(c_{xz,\mathbf{k}\alpha}^{\dagger}\boldsymbol{\sigma}{}_{\alpha\beta}c_{xz,\mathbf{k}+\mathbf{Q}_{2}\beta}^{\phantom{\dagger}}+\mathrm{h.c.}\right)\label{H_SDW_aux}
\end{align}
which, transformed to the spinor representation, yields: 
\begin{align}
\mathcal{H}_{\mathrm{SDW}}=\mathbf{M}_{1}\cdot\sum_{\mathbf{k}}\left[\psi_{\Gamma,\mathbf{k}}^{\dagger}\left(\tilde{\tau}^{1}\otimes\boldsymbol{\sigma}\right)\psi_{X,\mathbf{k}+\mathbf{Q}_{1}}^{\phantom{\dagger}}+\mathrm{h.c.}\right]\nonumber \\
+\mathbf{M}_{2}\cdot\sum_{\mathbf{k}}\left[\psi_{\Gamma,\mathbf{k}}^{\dagger}\left(-\tau^{-}\otimes\boldsymbol{\sigma}\right)\psi_{Y,\mathbf{k}+\mathbf{Q}_{2}}^{\phantom{\dagger}}+\mathrm{h.c.}\right]\label{H_SDW}
\end{align}

Note that, in the language of Ref. \cite{cvetkovic01}, the SDW Hamiltonian
transforms under the $E_{M_{4}}$ two-dimensional irreducible representation
of $P4/nmm$.

\section{Anisotropic magnetic free energy}

\label{sec:second_order_free_energy}

To understand the origin of the magnetic anisotropies, we first review
the group theoretical arguments of Ref. \cite{cvetkovic01}. In the
absence of SOC, all the components of the magnetic order parameters
belong to the same irreducible representation $E_{M_{4}}$, as shown
above, and the free energy depends only on the invariant form $\mbf{M}_{1}^{2}+\mbf{M}_{2}^{2}$.
However, with the introduction of SOC, the spin and orbital degrees
of freedom are no longer independent. Consequently, the components
of $\mbf{M}_{i}$ must belong to different irreducible representations
of $P4/nmm$, if one enforces the combination of the spin \textit{and}
orbital parts of the magnetic order parameter to still transform under
$E_{M_{4}}$. As a result, the individual components of $\mbf{M}_{1}$
and $\mbf{M}_{2}$ transform according to the following two-dimensional
irreducible representations \cite{cvetkovic01} 
\begin{eqnarray}
E_{M_{1}}\colon\begin{pmatrix}M_{1,x}\\
M_{2,y}
\end{pmatrix}\quad E_{M_{2}}\colon\begin{pmatrix}M_{1,y}\\
M_{2,x}
\end{pmatrix}\quad E_{M_{3}}\colon\begin{pmatrix}M_{1,z}\\
M_{2,z}
\end{pmatrix}\,.\label{eq:magnetic_irreps}
\end{eqnarray}

Because these components belong to different irreducible representations,
they will, in general, have different transition temperatures. Therefore,
the free energy must acquire the leading-order anisotropic terms:
\begin{figure}[t]
\centering \includegraphics[width=0.45\textwidth]{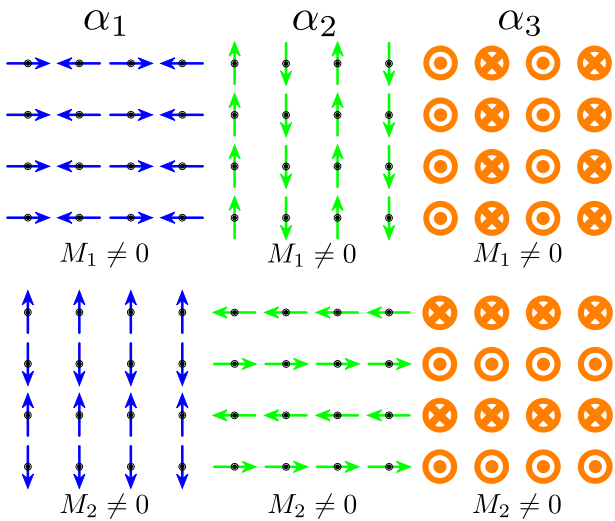}
\protect\protect\protect\protect\caption{\label{fig:magnetic_orders_C2} Sketch of the different uniaxial (i.e.
stripe-like) magnetic configurations corresponding to the different
anisotropic terms in the magnetic free energy (\ref{eq:free_energy_second_order})
with coefficients $\alpha_{1}$, $\alpha_{2}$, and $\alpha_{3}$.}
\end{figure}

\begin{eqnarray}
\delta F & = & \alpha_{1}\left(M_{1,x}^{2}+M_{2,y}^{2}\right)\nonumber \\
 &  & +\alpha_{2}\left(M_{1,y}^{2}+M_{2,x}^{2}\right)\nonumber \\
 &  & +\alpha_{3}\left(M_{1,z}^{2}+M_{2,z}^{2}\right)\,.\label{eq:free_energy_second_order}
\end{eqnarray}
The smallest $\alpha_{i}$ coefficient determines which type of magnetic
order condenses first. In Fig. \ref{fig:magnetic_orders_C2}, we show
separately the real-space spin configurations corresponding to the
components of $\mathbf{M}_{1}$ and $\mathbf{M}_{2}$ associated with
each coefficient $\alpha_{i}$. Specifically, if $\alpha_{1}<\alpha_{2},\alpha_{3}$,
then $\mathbf{M}_{i}$ points parallel to the ordering vector $\mathbf{Q}_{i}$
(of the unfolded zone); if $\alpha_{2}<\alpha_{1},\alpha_{3}$, $\mathbf{M}_{i}$
still points in-plane, but perpendicular to the ordering vector $\mathbf{Q}_{i}$.
Finally, if $\alpha_{3}<\alpha_{1},\alpha_{2}$, $\mathbf{M}_{i}$
points out-of-plane. Note that this analysis does not reveal whether
only either $\mathbf{M}_{1}$ or $\mathbf{M}_{2}$ condense, or if
both condense simultaneously. To establish the actual ground state,
it is necessary to go to higher order in the free energy. We will
come back to this point in Section \ref{sec:c2vsc4}. Note
that the spin-anisotropic terms preserve the tetragonal symmetry of
the system.

\begin{figure}[t]
\centering \includegraphics[width=0.5\textwidth]{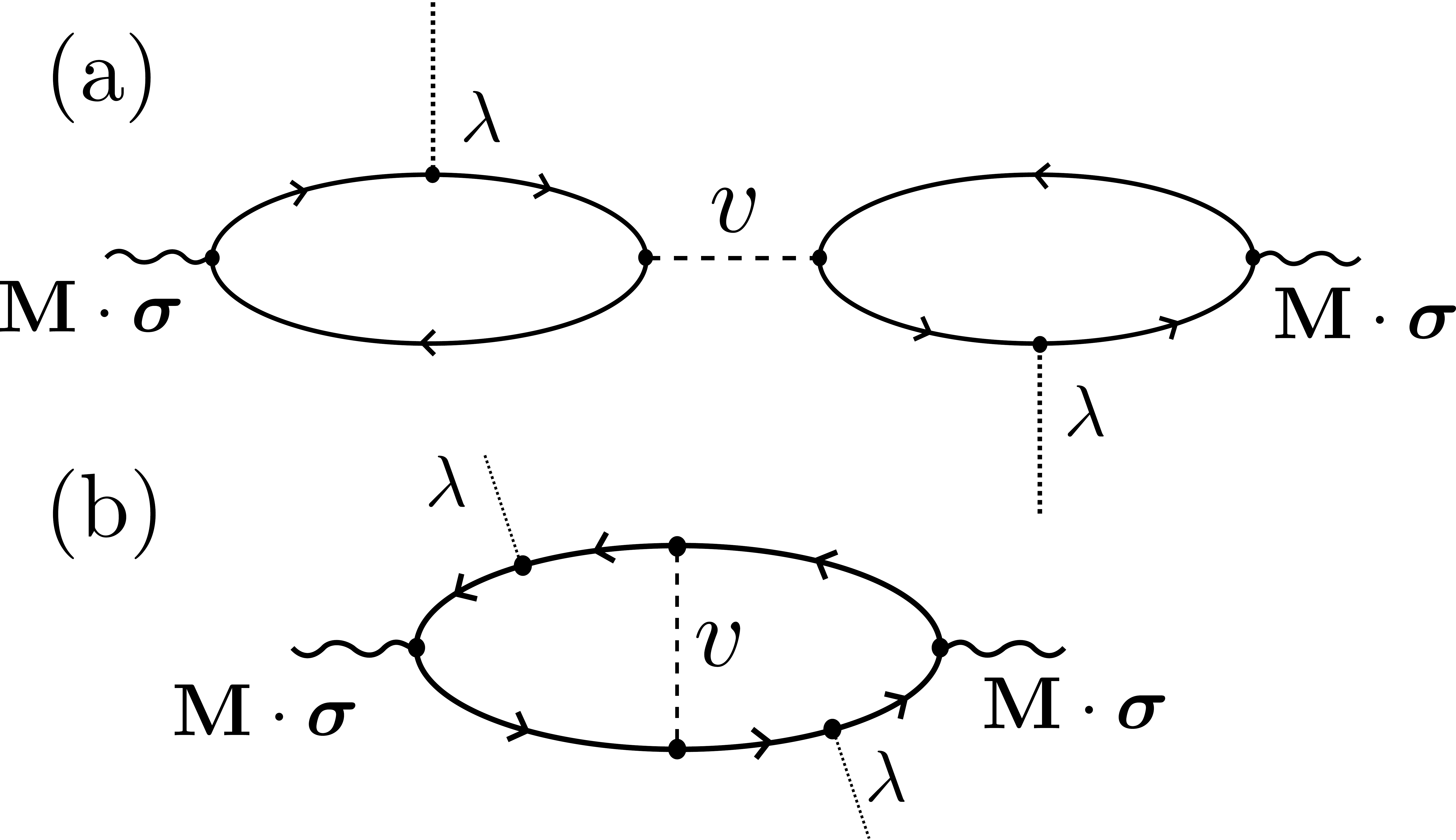}
\protect\protect\protect\protect\caption{\label{fig:diagram_examples} Schematic representation of the two
distinct type of Feynman diagrams at $\mathcal{O}(v,\lambda^{2})$:
two-loop diagrams (a) and one-loop diagrams (b). Only the one-loop
diagrams contribute to the anisotropic terms.}
\end{figure}

Here, our goal is to evaluate microscopically the $\alpha_{i}$ coefficients
using the model of the previous section. Within the non-interacting
part of the model, we find that even the presence of spin-orbit coupling
does not introduce magnetic anisotropies. The reason is that the model
effectively has an enlarged $P4/nmm\otimes P4/nmm$ symmetry, since
the states at $\Gamma$ and the states at $M$ are treated independently.
Of course, the fact that these states are connected in realistic tight-binding
models ensures that some level of spin-anisotropy will be introduced
at the non-interacting level. Such an effect will be likely a high-energy
effect, as it involves states away from the Fermi level \cite{Ahn}.
Here, instead, we focus on the low-energy contributions to the spin
anisotropy. Consequently, they must come from interactions -- particularly,
from the interaction terms that couple the states at $\Gamma$ and
at $M$, and therefore remove the enlarged $P4/nmm\otimes P4/nmm$
symmetry. These are precisely the terms listed in Eq. (\ref{eq:hamilton_interaction}).

We proceed with a straightforward diagrammatic approach by dressing
the non-interacting particle-hole bubble with the SOC term $\lambda$
in Eq. (\ref{H_SOC}) and with the interactions $v_{i}$ in Eq. (\ref{eq:hamilton_interaction}).
The SDW vertices coupling the magnetic order parameters to the non-interacting
Green's functions are those derived in Eq. (\ref{H_SDW}). Both $\lambda$
and $v_{i}$ are treated perturbatively to leading order. Because
terms of the order $\mathcal{O}(\lambda)$ are forbidden by symmetry,
we consider the diagrams of the orders $\mathcal{O}(\lambda^{2})$
and $\mathcal{O}(v_{i})$.

To order $\mathcal{O}(v_{i})$, there are two distinct types of interaction-dressed
diagrams, as depicted in Fig. \ref{fig:diagram_examples}. On top
of that, to order $\mathcal{O}(\lambda^{2})$, each of the two diagrams
can be dressed by a pair of SOC legs in eight different ways. Because
symmetry requirements forbid terms that couple directly $\mathbf{M}_{1}$
and $\mathbf{M}_{2}$ at the quadratic level, the pair of SOC legs
must correspond to the same SOC term, i.e. either $h_{\Gamma}^{\text{SOC}}(\mbf{k})$
or $h_{M}^{\text{SOC}}(\mbf{k})$ in Eq. (\ref{H_SOC}). Explicit
calculation of the traces over the Pauli-matrices reveals also that
the only combinations of SOC legs that yield anisotropic magnetic
terms are those in which one SOC leg appears in the upper-right (lower-right)
part of the diagram and the other SOC leg appears in the lower-left
(upper-left) part of the diagram.

\begin{figure*}[t]
\centering \includegraphics[width=0.95\textwidth]{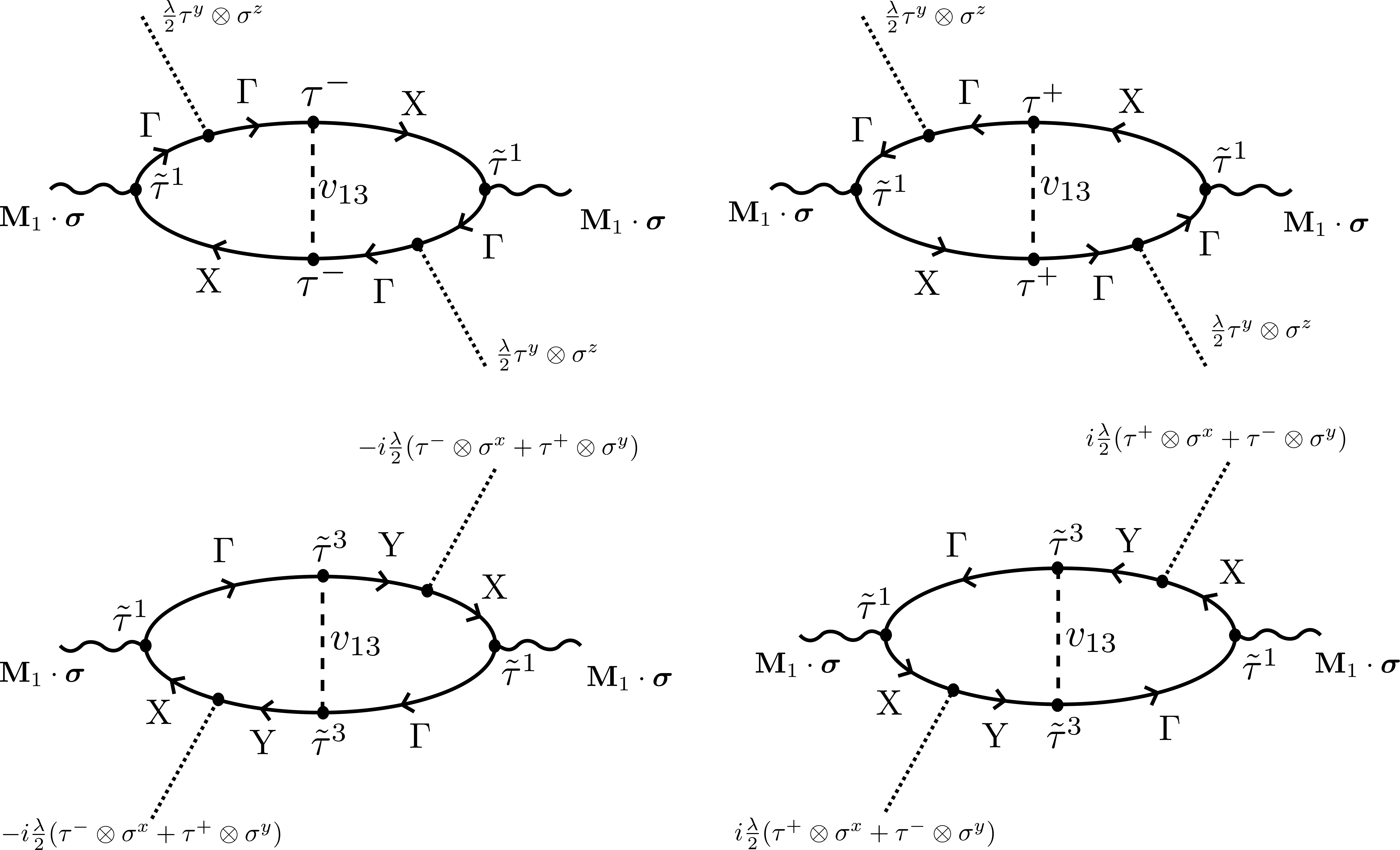}
\protect\protect\protect\protect\caption{\label{fig:second_order_g13} Illustration of distinct second-order
one-loop diagrams in the case where the electron-electron interaction
is given by $v_{13}$ and the magnetic order parameter is $\mbf{M}_{1}$.
Note that the electron-electron vertex depends on the direction of
momentum, i.e. the upper left and upper right diagrams are not identical.}
\end{figure*}

We find that all two-loop diagrams (i.e. those represented in Fig.
\ref{fig:diagram_examples}a) vanish, and therefore do not contribute
to the magnetic anisotropy term (\ref{eq:free_energy_second_order}).
We show this explicitly in Appendix \ref{app:two_loop_diagrams}.
Therefore, all that is left is to compute the one-loop diagram represented
in Fig. \ref{fig:diagram_examples}b. The calculation is tedious but
straightforward. To illustrate it, consider the interaction $v_{13}$.
The one-loop diagrams contributing to the anisotropic terms $M_{1,\mu}^{2}$
are shown in Fig. \ref{fig:second_order_g13}. As mentioned above,
there are two possible placements for the two SOC legs, in opposite
sides of the loop. For each diagram, one has to also consider its
hermitian-conjugate partner, since the interaction vertices and the
SDW vertices are not adjoint operators. Furthermore, the diagrammatic
rules derived for this problem impose an overall minus sign to each
one-loop diagram, and enforce the trace over the Pauli matrices to
be taken in the direction opposite to the arrows. Finally, to compute
these traces, it is useful to employ the following Pauli matrix identity:

\begin{equation}
\mathrm{tr}\left(\sigma^{i}\sigma^{j}\sigma^{k}\sigma^{l}\right)=2\left(\delta^{ij}\delta^{kl}-\delta^{ik}\delta^{jl}+\delta^{il}\delta^{jk}\right)\label{eq_Pauli}
\end{equation}

A straightforward evaluation of these four diagrams gives then three
different anisotropic terms for the magnetic free energy: 
\begin{widetext}
\begin{eqnarray}
\delta F & =- & v_{13}\frac{\lambda^{2}}{2}\sum_{\substack{\mbf{kk'}\\
\omega_{n}\omega_{n'}
}
}\Big(\text{tr}\left(\tilde{\tau}^{1}G_{X}'\tau^{-}G_{\Gamma}'\tau^{y}G_{\Gamma}'\tilde{\tau}^{1}G_{X}\tau^{-}G_{\Gamma}\tau^{y}G_{\Gamma}\right)+\text{tr}\left(\tilde{\tau}^{1}G_{\Gamma}'\tau^{y}G_{\Gamma}'\tau_{+}G_{X}'\tilde{\tau}^{1}G_{\Gamma}\tau^{y}G_{\Gamma}\tau^{+}G_{X}\right)\Big)\left(M_{1,z}^{2}-M_{1,x}^{2}-M_{1,y}^{2}\right)\nonumber \\
 & + & v_{13}\frac{\lambda^{2}}{2}\sum_{\substack{\mbf{kk'}\\
\omega_{n}\omega_{n'}
}
}\Big(\text{tr}\left(\tilde{\tau}^{1}G_{X}'\tau^{-}G_{Y}'\tilde{\tau}^{3}G_{\Gamma}'\tilde{\tau}^{1}G_{X}\tau^{-}G_{Y}\tilde{\tau}^{3}G_{\Gamma}\right)+\text{tr}\left(\tilde{\tau}^{1}G_{\Gamma}'\tilde{\tau}^{3}G_{Y}'\tau^{+}G_{X}'\tilde{\tau}^{1}G_{\Gamma}\tilde{\tau}^{3}G_{Y}\tau^{+}G_{X}\right)\Big)\left(M_{1,x}^{2}-M_{1,y}^{2}-M_{1,z}^{2}\right)\nonumber \\
 & + & v_{13}\frac{\lambda^{2}}{2}\sum_{\substack{\mbf{kk'}\\
\omega_{n}\omega_{n'}
}
}\Big(\text{tr}\left(\tilde{\tau}^{1}G_{X}'\tau^{+}G_{Y}'\tilde{\tau}^{3}G_{\Gamma}'\tilde{\tau}^{1}G_{X}\tau^{+}G_{Y}\tilde{\tau}^{3}G_{\Gamma}\right)+\text{tr}\left(\tilde{\tau}^{1}G_{\Gamma}'\tilde{\tau}^{3}G_{Y}'\tau^{-}G_{X}'\tilde{\tau}^{1}G_{\Gamma}\tilde{\tau}^{3}G_{Y}\tau^{-}G_{X}\right)\Big)\left(M_{1,y}^{2}-M_{1,x}^{2}-M_{1,z}^{2}\right)\nonumber \\
\delta F & = & v_{13}\lambda^{2}\Big(\sum_{\mbf{k},\omega_{n}}[G_{X}]_{12}[G_{Y}]_{12}[G_{\Gamma}]_{21}\Big)^{2}\left(M_{1,x}^{2}-M_{1,y}^{2}-M_{1,z}^{2}\right)\,.\label{eq_calc}
\end{eqnarray}

\end{widetext}

Here, the primed Green's functions are short-handed notations for
$G'_{i}=G_{i}\left(\mathbf{k}',\omega'_{n}\right)$. The Green functions
are given by 
\begin{eqnarray}
[G_{A}]_{ij}=\sum_{m}\frac{a_{m}^{i}(\mbf{k})_{A}a_{m}^{j}(\mbf{k})_{A}^{\ast}}{i\omega_{n}-(\epsilon_{A,m}(\mbf{k})-\mu)}\,,
\end{eqnarray}
where $\epsilon_{A,m}(\mbf{k})$ are the eigenenergies of the matrix
$h_{A}(\mbf{k})$ and $a_{m}^{i}(\mbf{k})_{A}$ is the unitary transformation
between the spinor basis and the band basis. In Eq. (\ref{eq_calc}),
the coefficient of the term $\left(M_{1,z}^{2}-M_{1,x}^{2}-M_{1,y}^{2}\right)$
vanishes because of the antisymmetry of the $G_{\Gamma}\tau^{y}G_{\Gamma}$
matrix in spinor space, whereas the coefficient of $\left(M_{1,y}^{2}-M_{1,x}^{2}-M_{1,z}^{2}\right)$
vanishes because $\left[G_{\Gamma}\right]_{12}\left(\mathbf{k}\right)\propto k_{x}k_{y}$,
causing the sum over momentum to vanish.

Note that while the upper diagrams shown in Fig. \ref{fig:second_order_g13}
introduce the anisotropy between the in-plane and out-of-plane components
of the magnetic order parameter, the lower diagrams contribute to
the anisotropy between $M_{1,x}$ and $M_{1,y}$. The reason for this
is the character of the SOC in the effective model, which remains
diagonal in the spin sector near the $\Gamma$-point due to $xz$-
and $yz$-orbital characters of the electronic states. As a result,
particle-hole excitations involving fermions from the $\Gamma-$point
and from the $X$/$Y$-point do not allow for a spin-flip, which would
be necessary to generate the anisotropy between $M_{1,x}$ and $M_{1,y}$.
Only the inclusion of the particle-hole excitations between the two
electron pockets, as described by the lower diagrams, generates the
anisotropy between the $x$ and $y$ components of the magnetization.

Repeating the same calculation for $\mathbf{M}_{2}$ gives the same
result, but with $M_{1,x}\rightarrow M_{2,y}$ and $M_{1,y}\rightarrow M_{2,x}$,
as expected by symmetry. Therefore, we can recast this contribution
to the free energy in the form of Eq. (\ref{eq:free_energy_second_order})
via differences in the anisotropic coefficients $\alpha$: 
\begin{eqnarray}
\alpha_{2}^{(v_{13})}-\alpha_{1}^{(v_{13})} & = & -2v_{13}\lambda_{M}^{2}\Big(\sum_{\mbf{k},\omega_{n}}[G_{X}]_{12}[G_{Y}]_{12}[G_{\Gamma}]_{21}\Big)^{2}\,\\
\alpha_{3}^{(v_{13})}-\alpha_{1}^{(v_{13})} & = & 0\,.
\end{eqnarray}
The same procedure applied to the other interactions $v_{15}$, $v_{17}$,
and $v_{19}$ reveals that only the first two give rise to anisotropic
terms. The final result for the anisotropic coefficients is: 
\begin{eqnarray}
\alpha_{2}-\alpha_{1} & = & 2v_{15}\lambda^{2}\Big(\sum_{\mbf{k},\omega_{n}}[G_{\Gamma}]_{11}[G_{X}]_{11}[G_{Y}]_{22}\Big)^{2}\nonumber \\
 &  & -2v_{13}\lambda^{2}\Big(\sum_{\mbf{k},\omega_{n}}[G_{\Gamma}]_{21}[G_{X}]_{12}[G_{Y}]_{12}\Big)^{2}\,,\label{eq:a2a1_expression}\\
\alpha_{3}-\alpha_{1} & = & 2v_{17}\lambda^{2}\Big(\sum_{\mbf{k},\omega_{n}}[G_{\Gamma}i\tau^{y}G_{\Gamma}]_{12}[G_{X}]_{11}\Big)^{2}\nonumber \\
 &  & -2v_{13}\lambda^{2}\Big(\sum_{\mbf{k},\omega_{n}}[G_{\Gamma}]_{21}[G_{X}]_{12}[G_{Y}]_{12}\Big)^{2}\,.\label{eq:a3a1_expression}
\end{eqnarray}
Mapping these interactions back to the more familiar multi-orbital
Hubbard model, as done in Ref. \cite{cvetkovic01}, reveals that $v_{13}=v_{15}=v_{17}=J$,
while the non-contributing term is $v_{19}=U/2$. Thus, our microscopic
calculation reveals that the low-energy magnetic anisotropy arises
from a combination of the SOC and of the Hund's rule coupling. This
anisotropy is present in the paramagnetic tetragonal phase, and does
not require orbital or nematic order.

\begin{figure}
\centering \includegraphics[width=0.45\textwidth]{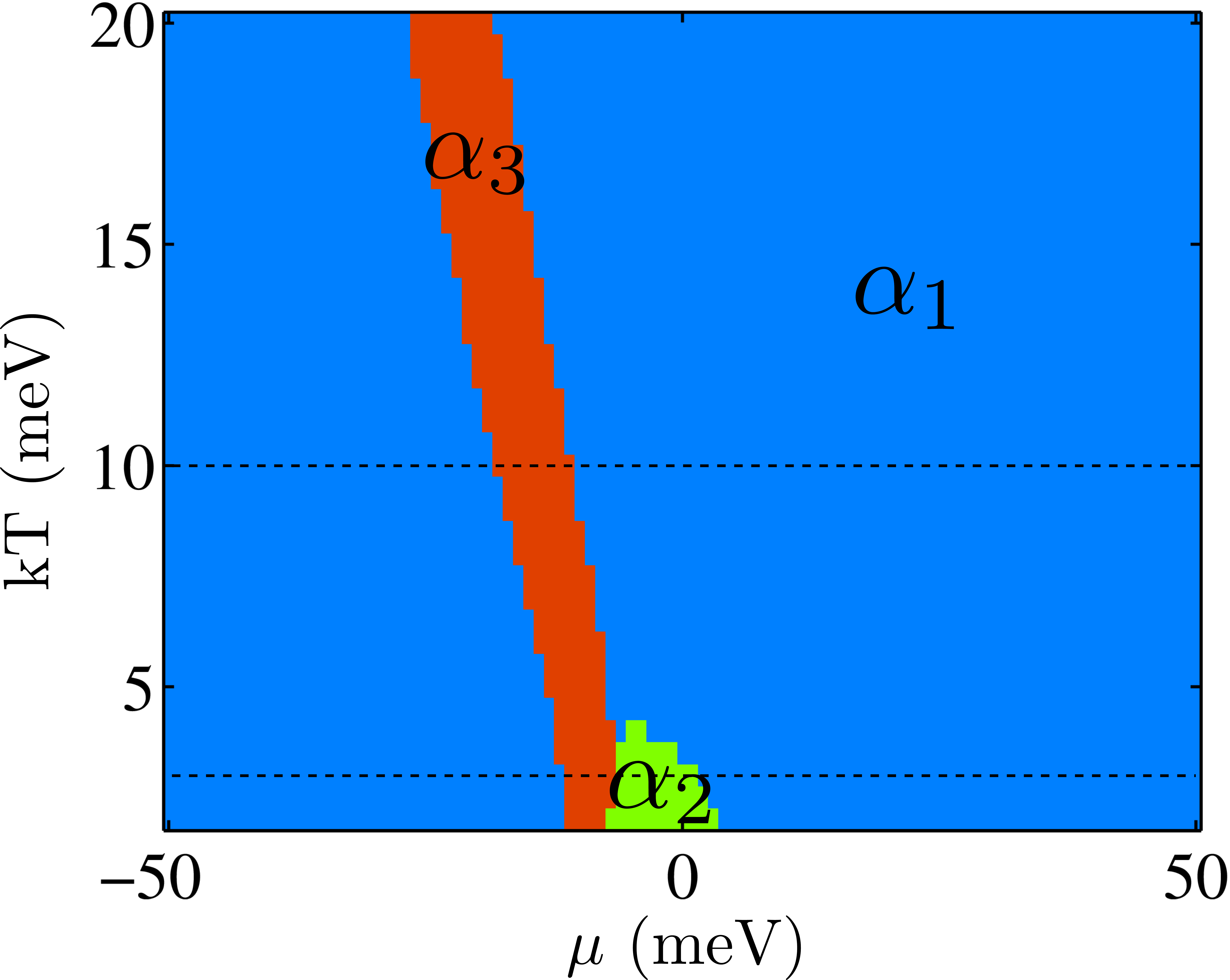}
\protect\protect\protect\protect\caption{\label{fig:XYZ_phase_diagram} Doping-temperature phase diagram displaying
the smallest magnetic anisotropy coefficient $\alpha$. When $\alpha_{1}$
is the smallest, the moments point in-plane and parallel to the ordering
vectors; when $\alpha_{2}$ is the smallest, the moments point in-plane
but perpendicular to the ordering vectors; finally, when $\alpha_{3}$
is the smallest, the moments point out-of-plane. The corresponding
uniaxial configurations are shown in Fig. \ref{fig:magnetic_orders_C2}.
Note that temperature here actually refers to the magnetic transition
temperature, as our model approaches the onset of long-range magnetic
order from the paramagnetic state. }
\end{figure}

\begin{figure}
\centering \includegraphics[width=0.45\textwidth]{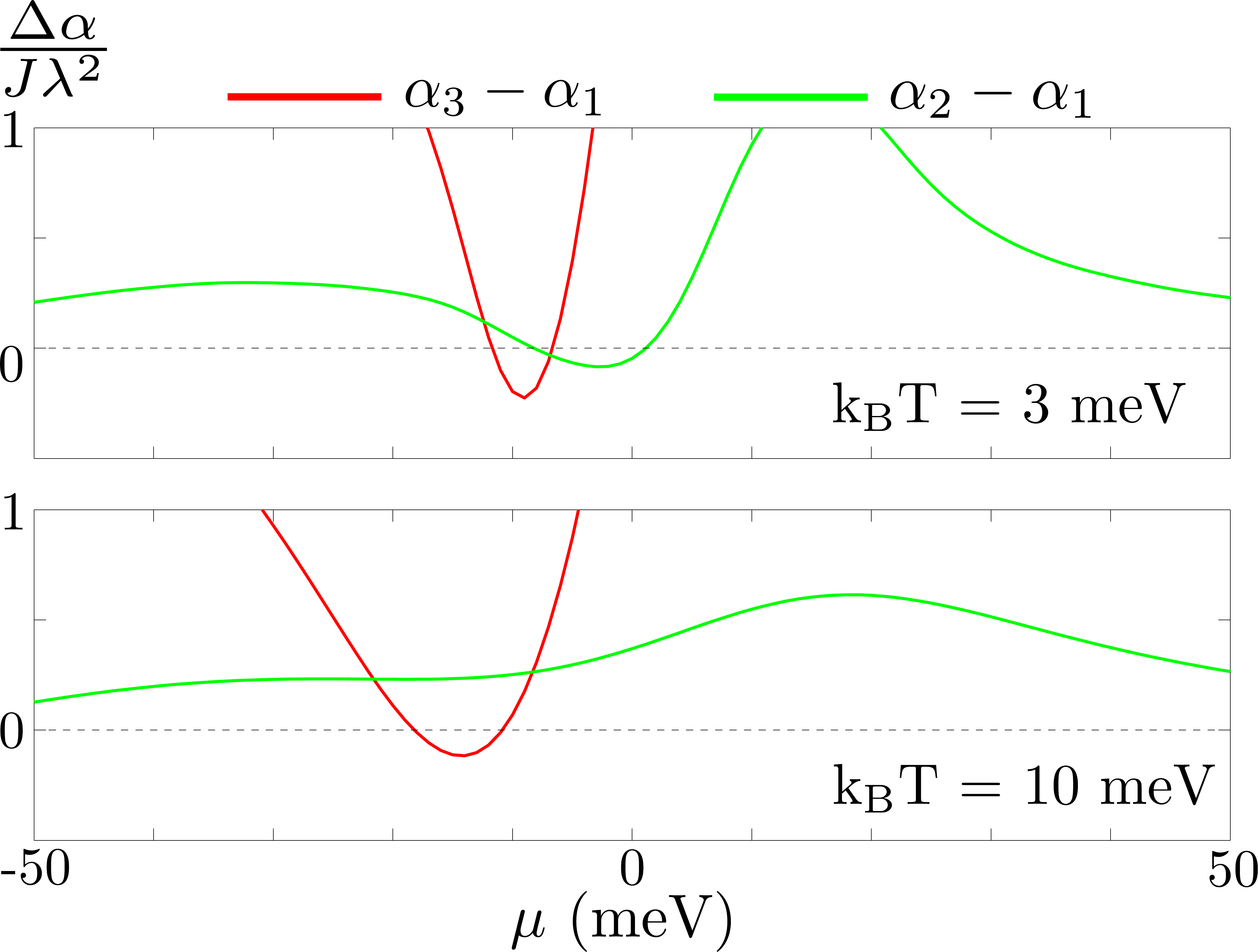}
\protect\protect\protect\caption{\label{fig:alpha_cut} Plots showing the dependence of $\alpha_{3}-\alpha_{1}$
(red curve) and $\alpha_{2}-\alpha_{1}$ (green curve) for two distinct
temperatures (see the dashed lines in Fig. \ref{fig:XYZ_phase_diagram})
as function of the chemical potential. }
\end{figure}

It is now straightforward to determine which of the three terms in
the free energy Eq. (\ref{eq:free_energy_second_order}) dominates.
For instance, if both $\alpha_{2}-\alpha_{1}>0$ and $\alpha_{3}-\alpha_{1}>0$,
$\alpha_{1}$ is the smallest of the three coefficients and the ordered
components of the magnetic moments will be $M_{1,x}$ and/or $M_{2,y}$.
By evaluating the expressions in Eqs. (\ref{eq:a2a1_expression})
and (\ref{eq:a3a1_expression}) numerically, using the parameters
that give the band dispersions and Fermi surface of Fig. \ref{fig:band_structure},
we can establish an effective ``doping-temperature phase diagram''
for the dominant anisotropy term as function of different values of
the chemical potential $\mu$ and of the magnetic transition temperature
$T$. Because the phase boundaries of this phase diagram are given by the conditions $\alpha_2 = \alpha_1$ or $\alpha_3 = \alpha_2$, and because these coefficients are independent of $U$ and have $J$ as an overall pre-factor (see Eqs. (36) and (37)), the phase boundaries do not change by varying $U$ and $J$. The phase diagram, shown in fig. \ref{fig:XYZ_phase_diagram},
reveals that for most of the parameter space considered here, the
$\alpha_{1}$ term is the smallest one, implying that the magnetic
moments point parallel to their ordering vectors $\mathbf{Q}_{i}$
below the magnetic transition. There is a small range of parameters
in which the moments lie in-plane, but perpendicular to their ordering
vectors (i.e. $\alpha_{2}$ is the smallest). Such a parameter regime
is likely not relevant for the iron pnictides, since it would require
an ``undoped'' composition (i.e. $\mu=0$) to display a rather small
magnetic transition temperature. Most interestingly, we find a robust
region in which the moments point out-of-plane (i.e. $\alpha_{3}$
is the smallest). This happens at any temperature, but always in the
hole-doped side of the phase diagram ($\mu<0$). In Fig. \ref{fig:alpha_cut},
we plot a zoom of the behavior of $\alpha_{2}-\alpha_{1}$ and $\alpha_{3}-\alpha_{1}$
as function of the chemical potential for two fixed temperatures to
illustrate the different regimes obtained. Note that, for most of
the phase diagram, $\left(\alpha_{2}-\alpha_{1}\right)\ll\left(\alpha_{3}-\alpha_{1}\right)$,
regardless of the value of $J\lambda^{2}$, which
appears as an overall prefactor of all $\alpha_{i}$ terms. This
implies that the spin anisotropy behaves effectively as an easy-plane
anisotropy. The fact that $\alpha_{3}$ becomes the
smallest coefficient in a narrow region of the phase diagram can be
attributed to the fact that the term $\sum_{\mbf{k},\omega_{n}}[G_{\Gamma}i\tau^{y}G_{\Gamma}]_{12}[G_{X}]_{11}$
in Eq. (\ref{eq:a3a1_expression}) changes sign from hole-doping to
electron-doping. This behavior can be understood qualitatively by
considering a hypotethical band structure in which all pockets are
perfectly nested, $G_{\Gamma}=\left(i\omega_{n}+\varepsilon\right)\otimes\tau^{0}$
and $G_{X}=G_{Y}=\left(i\omega_{n}-\varepsilon\right)\otimes\tau^{0}$. A straightforward calculation reveals that the three-Green's
function term above has different signs for $\mu>0$ and $\mu<0$, implying
that it must vanish for a certain chemical potential value in the
case of a realistic band structure.

Our results reveal not only an important asymmetry between electron-
and hole-doping, but also agree qualitatively with experiments. In
particular, neutron scattering measurements in hole- and electron-doped
BaFe$_{2}$As$_{2}$ \cite{waser01,osborn01,allred01,allred02} find
generally in-plane moments parallel to $\mathbf{Q}_{i}$ in the magnetically-ordered
state, except at a narrow hole-doping range in which the moments reorient
and point out-of-the-plane.

\section{tetragonal vs stripe magnetic order}

\label{sec:c2vsc4}

\begin{figure}
\centering \includegraphics[width=0.45\textwidth]{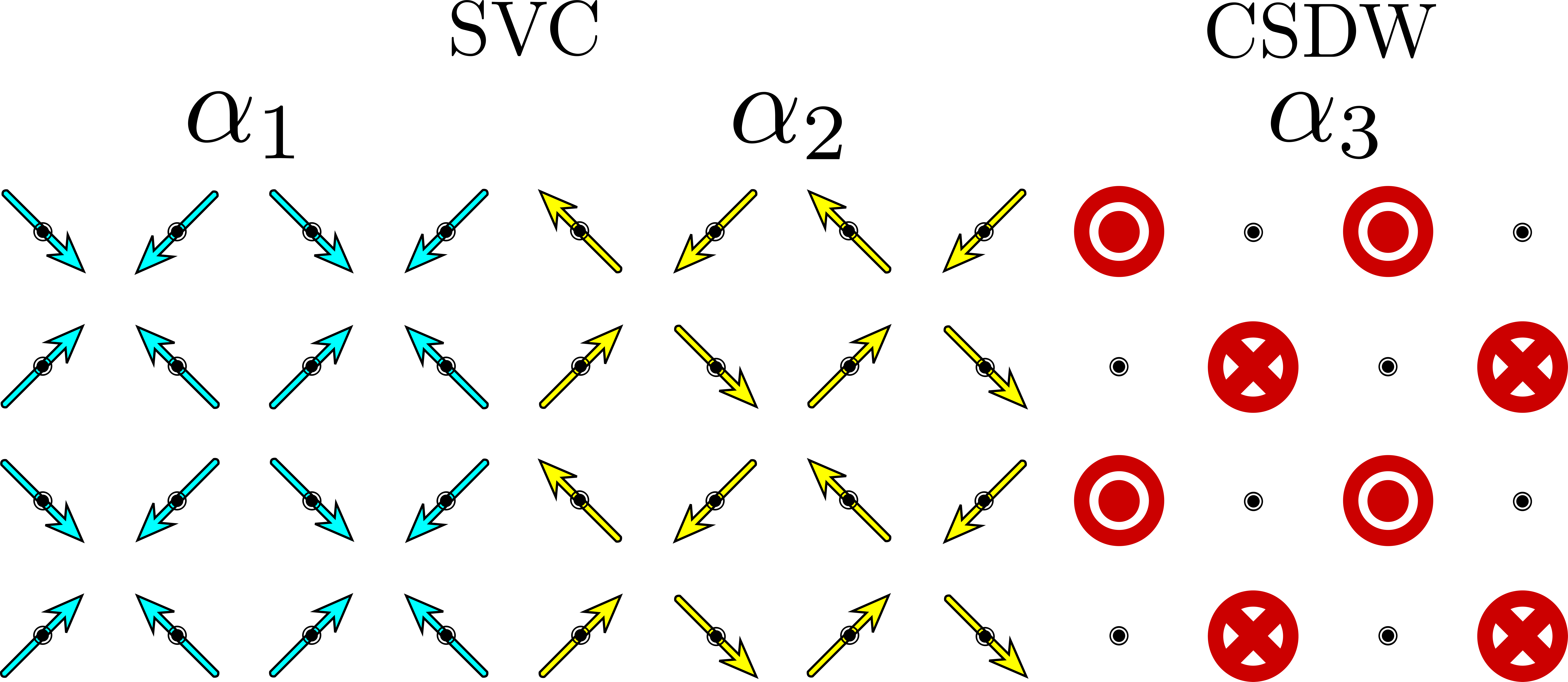}
\protect\protect\protect\protect\caption{\label{fig:magnetic_orders_C4} Sketch of the three possible biaxial
tetragonal magnetic phases. Here SVC is the spin vortex crystal phase
(with non-collinear magnetic moments) and CSDW is the charge-spin
density wave phase (with non-uniform magnetic moments).}
\end{figure}

The previous section established the direction of the magnetic moments,
but not the magnetic ground state. For instance, from the analysis
of the second-order terms of the free energy, it is impossible to
distinguish the cases in which either $\mathbf{M}_{1}$ or $\mathbf{M}_{2}$
condense (i.e. $\mbf{M}_{2}=0$ or $\mbf{M}_{1}=0$) from the case
in which both condense simultaneously ($\mbf{M}_{1}=\mbf{M}_{2}\neq0$).
The former case gives the striped orthorhombic magnetic phases shown
in Fig. \ref{fig:magnetic_orders_C2}, whereas the latter case gives
rise to a double-\textbf{Q }(i.e. biaxial) magnetic state that preserves
the tetragonal symmetry of the system. From the form of the anisotropic
terms in the free energy (see Sec. \ref{sec:second_order_free_energy}),
there are three different types of tetragonal magnetic ground states,
as shown in Fig. \ref{fig:magnetic_orders_C4}. Two of them correspond
to the so-called spin-vortex crystal phase (SVC), a non-collinear
state in which $M_{1,x}=M_{2,y}\neq0$ or $M_{1,y}=M_{2,x}\neq0$,
whereas the third one corresponds to the so-called charge-spin density-wave
phase (CSDW) \cite{fernandes05}, a non-uniform state in which $M_{1,z}=M_{2,z}\neq0$.

To determine whether the ground state corresponds to an orthorhombic
uniaxial SDW (i.e. either $\mbf{M}_{1}\neq0$ or $\mbf{M}_{2}\neq0$)
or to a tetragonal biaxial SDW (i.e. $\mbf{M}_{1}=\mbf{M}_{2}\neq0$),
we need to go to higher order in the free energy expansion. Symmetry
requires the free energy to have the form (in the absence of SOC):
\begin{eqnarray}
F^{(4)}(\mbf{M}_{1},\mbf{M}_{2}) & = & \frac{u}{2}\left(\mbf{M}_{1}^{2}+\mbf{M}_{2}^{2}\right)^{2}-\frac{g}{2}\left(\mbf{M}_{1}^{2}-\mbf{M}_{2}^{2}\right)^{2}\label{eqF_4_aux}\\
 & + & 2w(\mbf{M}_{1}\cdot\mbf{M}_{2})^{2}\,.
\end{eqnarray}

Minimizing this expression shows that the tetragonal biaxial state
is realized when $g<0$ or $g<-w$, whereas the orthorhombic uniaxial
state takes place when $g>0$ and $g>-w$.

The same model was derived by different itinerant approaches for the
magnetic instabilities of the iron pnictides, revealing different
parameter regimes in which the uniaxial or the biaxial states are
the ground states \cite{lorenzana01,eremin01,giovannetti01,wang01,kang01,gastiasoro01,brydon01}.
In this regard, the novelty of our approach relies on the relationship
between these ground states and the magnetic anisotropies, and also
on the employment of a low-energy model that respects all symmetries
of the FeAs plane, including the As puckering that enhances the size
of the Fe unit cell. Because $u$ and $g$ are non-zero even for vanishing
SOC and interactions, we compute only the contributions arising from
the non-interacting part of the Hamiltonian. This is achieved either
by standard diagrammatics or by explicitly integrating out the electronic
degrees of freedom. We find:

\begin{figure*}[t]
\centering \includegraphics[width=0.95\textwidth]{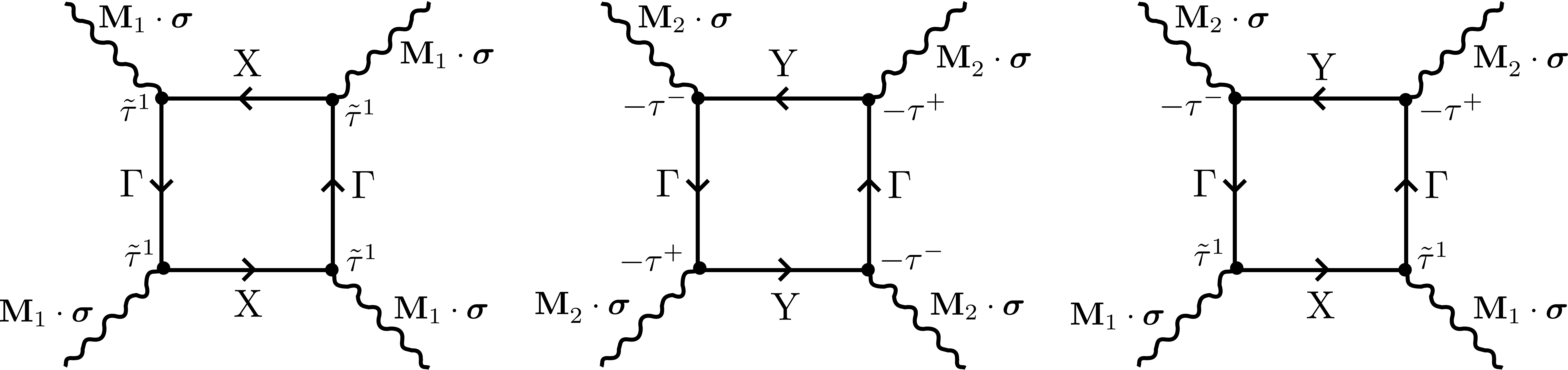}
\protect\protect\protect\protect\caption{\label{fig:fourth_order_diagrams} Illustration of the three distinct
fourth order diagrams that contribute to $F^{(4)}$. Note that the
right diagram has a symmetry factor of two, while the symmetry factor
of the middle and left diagrams is one. One might expect there to
be a fourth diagram with alternating $\mbf{M}_{1}$ and $\mbf{M}_{2}$,
however, the vertex coupling the SDW order parameter and the fermions
forbids it (see Appendix \ref{app:w_coeff} for more details).}
\end{figure*}

\begin{eqnarray}
F^{(4)} &  & =\mbf{M}_{1}^{4}\sum_{\mbf{k},\omega_{n}}\text{tr}\left(\tilde{\tau}^{1}G_{\Gamma}\tilde{\tau}^{1}G_{X}\tilde{\tau}^{1}G_{\Gamma}\tilde{\tau}^{1}G_{X}\right)\nonumber \\
 &  & \qquad+\mbf{M}_{2}^{4}\sum_{\mbf{k},\omega_{n}}\text{tr}\left(\tau^{+}G_{\Gamma}\tau^{-}G_{Y}\tau^{+}G_{\Gamma}\tau^{-}G_{Y}\right)\nonumber \\
 &  & \qquad+2\mbf{M}_{1}^{2}\mbf{M}_{2}^{2}\sum_{\mbf{k},\omega_{n}}\text{tr}\left(\tau^{+}G_{\Gamma}\tilde{\tau}^{1}G_{X}\tilde{\tau}^{1}G_{\Gamma}\tau^{-}G_{Y}\right)\nonumber \\
F^{(4)} &  & =\mbf{M}_{1}^{4}\sum_{\mbf{k},\omega_{n}}\left([G_{\Gamma}]_{11}^{2}[G_{X}]_{11}^{2}\right)\nonumber \\
 &  & \qquad+\mbf{M}_{2}^{4}\sum_{\mbf{k},\omega_{n}}\left([G_{\Gamma}]_{22}^{2}[G_{Y}]_{11}^{2}\right)\nonumber \\
 &  & \qquad+2\mbf{M}_{1}^{2}\mbf{M}_{2}^{2}\sum_{\mbf{k},\omega_{n}}\left([G_{\Gamma}]_{12}[G_{X}]_{11}[G_{Y}]_{11}\right)\,,\label{eq:sum_fourth_order_diagrams}
\end{eqnarray}
with the corresponding diagrams shown in Fig.~\ref{fig:fourth_order_diagrams}.
Rewriting the free energy in the form (\ref{eqF_4_aux}), we can readily
obtain $u$, $g$, and $w$: 
\begin{eqnarray}
u & = & \sum_{\mbf{k},\omega_{n}}\left([G_{\Gamma}]_{12}^{2}[G_{Y}]_{11}[G_{X}]_{11}+[G_{\Gamma}]_{11}^{2}[G_{X}]_{11}^{2}\right)\label{eq:u_expression}\\
g & = & \sum_{\mbf{k},\omega_{n}}\left([G_{\Gamma}]_{12}^{2}[G_{Y}]_{11}[G_{X}]_{11}-[G_{\Gamma}]_{11}^{2}[G_{X}]_{11}^{2}\right)\label{eq:g_expression}\\
w & = & 0\,.
\end{eqnarray}
One might expect the third diagram in Fig.~\ref{fig:fourth_order_diagrams}
to result in a non-vanishing $w$, however, contraction of the Pauli
matrices (see Eq.~\ref{eq_Pauli}) reveals that the $\mbf{M}_{1}\cdot\mbf{M}_{2}$
term cancels. More generally, $w=0$ is a robust property of our model,
a consequence of momentum conservation and the absence of a Fermi
pocket at $(\pi,\pi)$ (see Appendix~\ref{app:w_coeff}). Including
a hole-pocket at $(\pi,\pi)$ or interactions will however lead to
a non-zero contribution to $w$ \cite{wang01}.

\begin{figure}[b]
\centering \includegraphics[width=0.45\textwidth]{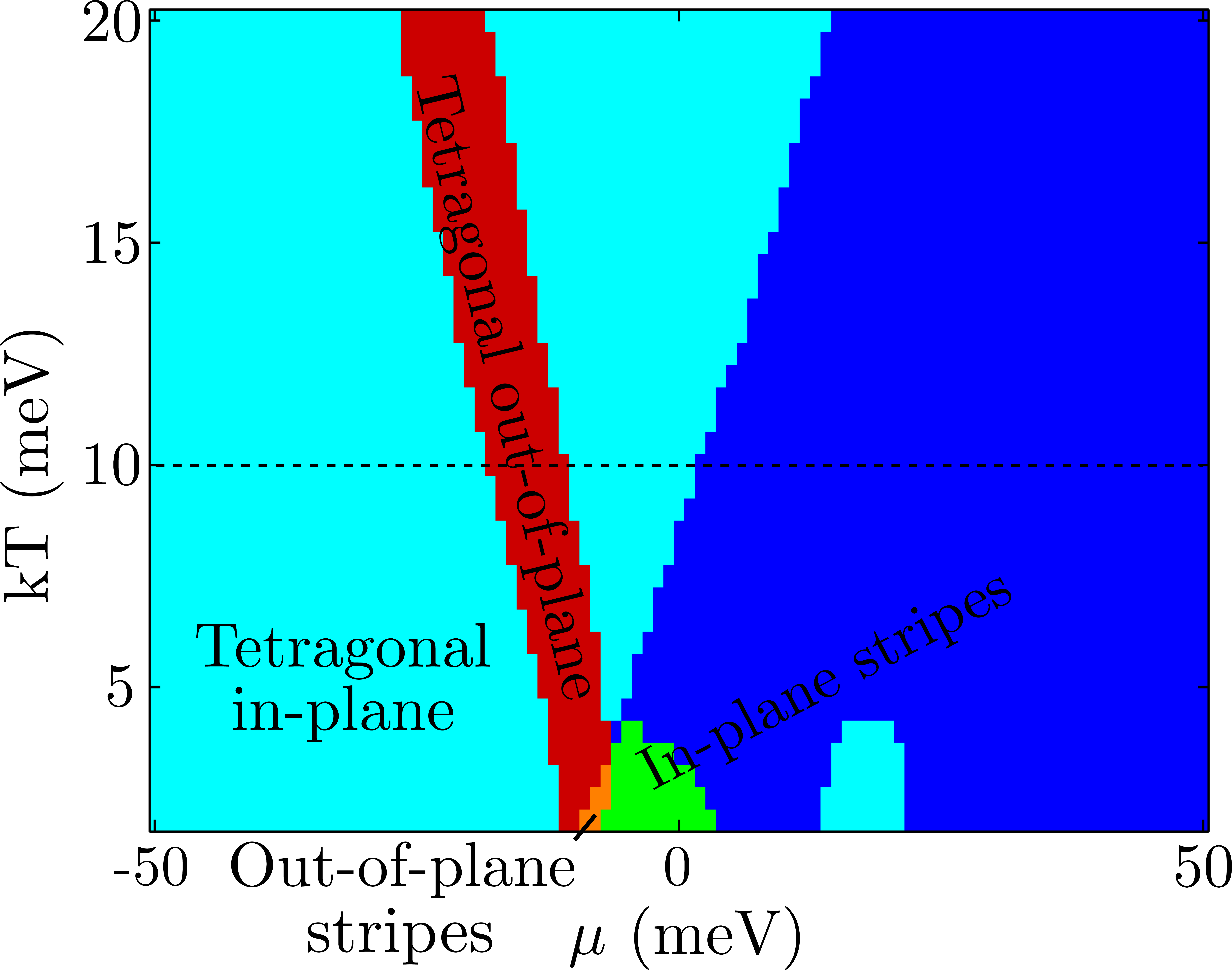}
\protect\protect\protect\protect\caption{\label{fig:phase_diagram} Doping-temperature phase diagram of the
different types of magnetic ground state (stripes or tetragonal) and
their corresponding spin orientation (in-plane or out-of-plane). The
color-code corresponds to the magnetic configurations shown in Figs.
\ref{fig:magnetic_orders_C2} and \ref{fig:magnetic_orders_C4}. Note
that temperature here actually refers to the magnetic transition temperature,
as our model approaches the onset of long-range magnetic order from
the paramagnetic state.}
\end{figure}

\begin{figure}
\centering \includegraphics[width=0.45\textwidth]{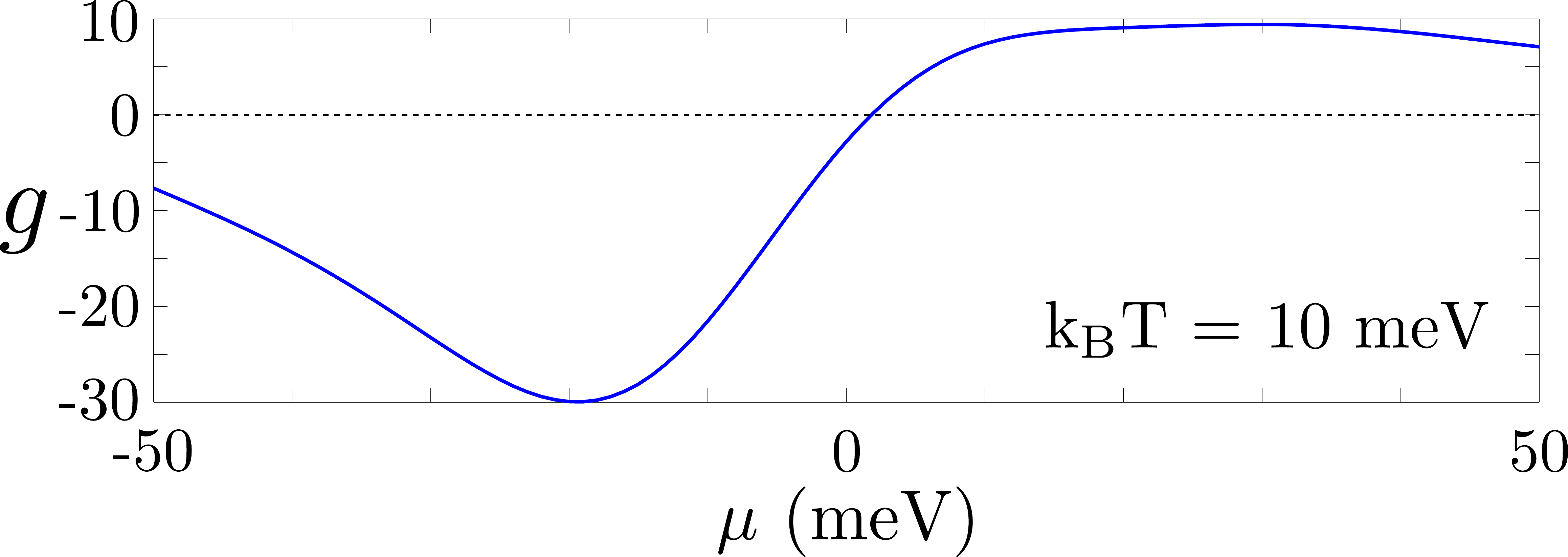} \protect\protect\protect\caption{\label{fig:g_cut} The quartic coefficient $g$ as a function of the
chemical potential $\mu$ for a constant temperature (see the dashed
line in Fig. \ref{fig:phase_diagram}). When $g$ is negative the
system chooses a tetragonal biaxial magnetic phase whereas for positive
$g$, the system selects an orthorhombic uniaxial stripe state.}
\end{figure}

We can now compute numerically the value of $g$ for the same ``doping-temperature
phase-diagram'' studied in the previous section, see Fig. \ref{fig:XYZ_phase_diagram}.
The combined result, shown in Fig. \ref{fig:phase_diagram}, accentuates
the asymmetry between hole- and electron-doping discussed previously.
In particular, while hole-doping tends to favor a tetragonal biaxial
SDW state, electron-doping tends to favor an orthorhombic uniaxial
SDW state. A cut with the behavior of $g$ as function of $\mu$ for
a fixed temperature is also shown in Fig. \ref{fig:g_cut}. To
gain more insight into the behavior of $g$, we consider once again
the hypothetical case of perfectly-nested bands, $G_{\Gamma}=\left(i\omega_{n}+\varepsilon\right)\otimes\tau^{0}$
and $G_{X}=G_{Y}=\left(i\omega_{n}-\varepsilon\right)\otimes\tau^{0}$.
In this case, from the equations above, $g<0$. Building
on the results of Ref. \cite{eremin01}, we expect that the sign of
$g$ will change once the two hole pockets become rather different
in size, such that one of them becomes poorly nested with the electron
pockets. Our calculations indicate that, for the general tight-binding
model studied here, this is favored by hole doping rather than electron
doping.

It is important to emphasize that these results should be understood
as general trends as function of the chemical potential, rather than
a full determination of the ground state for each specific value of
$\mu$. This is because, in contrast to the previous section, in which
the lowest order contribution to the spin anisotropy arises solely
from the SOC and the interactions, there are other potential contributions
to $g$ beyond the scope of the current work. Among these contributions,
we highlight the sizable magneto-elastic coupling, which should extend
the stripe phase to wider doping ranges \cite{shear_modulus,Paul11},
and interaction corrections, which can also favor the uniaxial over
the biaxial state \cite{eremin01,wang01}. With this word of caution,
we note that the tendency observed here that hole-doped compounds
are more favorable to a tetragonal magnetic phase as compared to their
electron-doped counterparts is in qualitative agreement with experiments,
which observe a small region of tetragonal magnetism near the optimally-hole
doped pnictides Ba$_{1-x}$Na$_{x}$Fe$_{2}$As$_{2}$ \cite{avci01},
Sr$_{1-x}$Na$_{x}$Fe$_{2}$As$_{2}$ \cite{allred02}, Ba$_{1-x}$K$_{x}$Fe$_{2}$As$_{2}$
\cite{bohmer01,mallett02}, and Ba(Fe$_{1-x}$Mn$_{x}$)$_{2}$As$_{2}$
\cite{kim01}.

Interestingly, in the first three compounds, neutron scattering has
shown that the onset of a tetragonal magnetic state takes place in
a region in which the magnetic moments reorient from in-plane to out-of-plane.
Within our analysis, this can be attributed to the robust region in
parameter space in which $\alpha_{3}$ is the smallest anisotropic
coefficient (see Fig. \ref{fig:phase_diagram}). More importantly,
this anisotropic coefficient removes the degeneracy between the two
types of tetragonal SDW phase -- the SVC and the CSDW states (see
Fig. \ref{fig:magnetic_orders_C4}) -- by favoring the latter. This
is expected to happen even if $w\neq0$, since the latter is a quartic
coefficient, whereas $\alpha_{3}$ is a quadratic coefficient. Therefore,
at least near the onset of the magnetic transition, it is the SOC
and the Hund's rule coupling that select the CSDW phase. Recently,
M{\"o}ssbauer \cite{allred02} and $\mu$SR \cite{mallet01} experiments
in Sr$_{1-x}$Na$_{x}$Fe$_{2}$As$_{2}$ and Ba$_{1-x}$K$_{x}$Fe$_{2}$As$_{2}$,
respectively, have reported direct evidence that indeed the CSDW state
is realized in the regime where the spin is reoriented and the magnetic
long-range order preserves tetragonal symmetry.

\section{Discussion and Conclusions}

\label{sec:conclusions}

In summary, we have shown that within a low-energy model that respects
the symmetries of the FeAs plane, magnetic anisotropy arises naturally
from the combination of the spin-orbit coupling and the Hund's rule
coupling. The magnetic anisotropy consists of three terms (see Sec.
\ref{sec:second_order_free_energy}). Although it cannot be mapped
generally on an easy-axis or an easy-plane term, it effectively behaves
as an easy-plane term for a large part of the parameter region studied
here, since $\left(\alpha_{2}-\alpha_{1}\right)\ll\left(\alpha_{3}-\alpha_{1}\right)$.
We found that, for most of the temperature-doping phase diagram, the
spin anisotropy is such that the magnetic moments point in-plane and
parallel to the direction of the ordering vector. For a small doping
range in the hole-doped side, across all temperatures studied, the
magnetic moments tend to reorient and point out-of-plane. These features
are consistent with those observed experimentally, including the spin
reorientation observed in the hole-doped pnictides Ba$_{1-x}$Na$_{x}$Fe$_{2}$As$_{2}$
\cite{waser01}, Sr$_{1-x}$Na$_{x}$Fe$_{2}$As$_{2}$ \cite{allred02},
Ba$_{1-x}$K$_{x}$Fe$_{2}$As$_{2}$ \cite{allred01}. We also found
a general tendency of tetragonal double-\textbf{Q} magnetic order
for the hole-doped side of the phase diagram, whereas the orthorhombic
single-\textbf{Q }stripe magnetic order is favored in the electron-doped
side. Although this is in general agreement with the experimental
observations in Ba$_{1-x}$Na$_{x}$Fe$_{2}$As$_{2}$ \cite{avci01},
Sr$_{1-x}$Na$_{x}$Fe$_{2}$As$_{2}$ \cite{allred02}, Ba$_{1-x}$K$_{x}$Fe$_{2}$As$_{2}$
\cite{bohmer01}, and Ba(Fe$_{1-x}$Mn$_{x}$)$_{2}$As$_{2}$ \cite{kim01},
our results seem to overestimate the size of the region in which the
tetragonal magnetic state is stable. A possible reason for this discrepancy
is that our model does not account for other factors that usually
favor the stripe over the tetragonal magnetic state, such as the magneto-elastic
coupling \cite{shear_modulus,Paul11} and the residual interactions
not directly responsible for the SDW instability \cite{eremin01,wang01}.
Yet, our results provide a clear connection between the spin reorientation
and the type of tetragonal magnetic state observed in the hole-doped
iron pnictides -- namely, the charge-spin density-wave state with
a non-uniform magnetization \cite{allred02,mallet01}.

An important consequence of our results is that the spin anisotropy
is not necessarily tied to the orbital order that is triggered across
the nematic/structural transition \cite{devereaux01}. Although it
is plausible that such an orbital order affects the spin anisotropy,
the latter exists already in the tetragonal paramagnetic state as
a result of the symmetry properties of the FeAs plane. In this regard,
it would be interesting to investigate how the spin anisotropies in
the tetragonal-paramagnetic phase studied here are connected to the
spin anisotropies in the low-temperature phase, after both magnetic
and nematic orders are well established. Finally, our results open
the important question of how this particular form of magnetic anisotropy
impacts the normal state properties of the iron pnictides. In particular,
the onset temperatures and the characters of the coupled nematic-magnetic
transitions are expected to be strongly affected by any form of spin
anisotropy \cite{Capati11,fernandes01,devereaux01}.

\section*{Acknowledgements}

We thank F. Ahn, M. Schulz, A. Chubukov, P. Dai, P. Orth, J. Knolle,
M. Korshunov, R. Osborn, P. Hirschfeld, Y. Zhao and O. Vafek for fruitful
discussions. MHC and BMA acknowledge financial support from a Lundbeckfond
fellowship (grant A9318). RMF and JK are supported by the U.S. Department
of Energy, Office of Science, Basic Energy Sciences, under award number
DE-SC0012336. The work of IE was supported by the Focus Program 1458
Eisen-Pniktide of the DFG, and by the German Academic Exchange Service
(DAAD PPP USA no. 57051534). IE also acknowledges the financial support of the Ministry of Education and Science of the Russian Federation in the framework of Increase Competitiveness Program of NUST MISiS (N 2-2014-015).

\appendix

\section{Expansion of tight-binding Hamiltonian for small $\mbf{k}$}

\label{app:heurestic_model}

In this appendix we derive explicitly the non-interacting Hamiltonian
$H_{0}$ introduced in sec. \ref{sec:model}. As explained in that
section, we need to project and expand the 5-orbital tight-binding
dispersion $\varepsilon_{\mu\nu}(\mbf{k})$, where $\mu,\nu=1,...,5$
are the orbital indices corresponding to $xz$, $yz$, $x^{2}-y^{2}$,
$xy$ and $3z^{2}-r^{2}$, respectively. The dispersions are given
by \cite{graser01} 
\begin{widetext}
\begin{eqnarray}
\varepsilon_{11} & = & \epsilon_{xz/yz}+2t_{x}^{11}\cos k_{x}+2t_{y}^{11}\cos k_{y}+4t_{xy}^{11}\cos k_{x}\cos k_{y}+2t_{xx/yy}^{11}\left(\cos2k_{x}-\cos2k_{y}\right)\nonumber \\
 & + & 4t_{xxy}^{11}\cos2k_{x}\cos k_{y}+4t_{xyy}^{11}\cos k_{x}\cos2k_{y}+4t_{xxyy}^{11}\cos2k_{x}\cos2k_{y}\,,\label{eq:M11}\\
\varepsilon_{22} & = & \epsilon_{xz/yz}+2t_{x}^{22}\cos k_{x}+2t_{y}^{22}\cos k_{y}+4t_{xy}^{22}\cos k_{x}\cos k_{y}-2t_{xx/yy}^{22}\left(\cos2k_{x}-\cos2k_{y}\right)\nonumber \\
 & + & 4t_{xxy}^{22}\cos2k_{x}\cos k_{y}+4t_{xyy}^{22}\cos k_{x}\cos2k_{y}+4t_{xxyy}^{22}\cos2k_{x}\cos2k_{y}\,,\\
\varepsilon_{33} & = & \epsilon_{x^{2}-y^{2}}+2t_{x/y}^{33}\left(\cos k_{x}+\cos k_{y}\right)+4t_{xy}^{33}\cos k_{x}\cos k_{y}2t_{xx/yy}^{33}\left(\cos2k_{x}+\cos2k_{y}\right)\,,\\
\varepsilon_{44} & = & \epsilon_{xy}+2t_{x/y}^{44}\left(\cos k_{x}+\cos k_{y}\right)+t_{xy}^{44}\cos k_{x}\cos k_{y}+2t_{xx/yy}^{44}\left(\cos2k_{x}+\cos2k_{y}\right)\nonumber \\
 & + & 4t_{xxy/xyy}^{44}\left(\cos2k_{x}\cos k_{y}+\cos k_{x}\cos2k_{y}\right)+4t_{xxyy}^{44}\cos2k_{x}\cos2k_{y}\,,\\
\varepsilon_{55} & = & \epsilon_{z^{2}}+2t_{x/y}^{55}\left(\cos k_{x}+\cos k_{y}\right)+2t_{xx/yy}^{55}\left(\cos2k_{x}\cos2k_{y}\right)\nonumber \\
 & + & 4t_{xxy/xyy}^{55}\left(\cos2k_{x}\cos k_{y}+\cos k_{x}\cos2k_{y}\right)+4t_{xxyy}^{55}\cos2k_{x}\cos2k_{y}\,,\\
\varepsilon_{12} & = & 4t_{xy}^{12}\sin k_{x}\sin k_{y}+4t_{xxy/xyy}^{12}\left(\sin2k_{x}\sin k_{y}+\sin k_{x}\sin2k_{y}\right)+4t_{xxyy}^{12}\sin2k_{x}\sin2k_{y}\,,\\
\varepsilon_{13} & = & i2t_{y}^{13}\sin k_{y}+i4t_{xy}^{13}\cos k_{x}\sin k_{y}-i4t_{xxy/xyy}^{13}\left(\cos k_{x}\sin2k_{y}-\cos2k_{x}\sin k_{y}\right)\,,\\
\varepsilon_{14} & = & i2t_{x}^{14}\sin k_{x}-i4t_{xy}^{14}\sin k_{x}\cos k_{y}+i4t_{xxy}^{14}\sin2k_{x}\cos k_{y}\,,\\
\varepsilon_{15} & = & i2t_{y}^{15}\sin k_{y}-i4t_{xy}^{15}\cos k_{x}\sin k_{y}-i4t_{xxyy}^{15}\cos2k_{x}\sin2k_{y}\,,\\
\varepsilon_{23} & = & i2t_{x}^{23}\sin k_{x}+i4t_{xy}^{23}\sin k_{x}\cos k_{y}-i4t_{xxy/xyy}^{23}\left(\sin2k_{x}\cos k_{y}-\sin k_{x}\cos2k_{y}\right)\,,\\
\varepsilon_{24} & = & -i2t_{y}^{24}\sin k_{y}+i4t_{xy}^{24}\cos k_{x}\sin k_{y}-i4t_{xyy}^{24}\cos k_{x}\sin2k_{y}\,,\\
\varepsilon_{25} & = & -i2t_{x}^{25}\sin k_{x}+i4t_{xy}^{25}\sin k_{x}\cos k_{y}+i4t_{xxyy}^{25}\sin2k_{x}\cos2k_{y}\,,\\
\varepsilon_{34} & = & 4t_{xxy/xyy}^{34}\left(\sin k_{x}\sin2k_{y}-\sin2k_{x}\sin k_{y}\right)\,,\\
\varepsilon_{35} & = & 2t_{x/y}^{35}\left(\cos k_{x}-\cos k_{y}\right)+4t_{xxy/xyy}^{35}\left(\cos2k_{x}\cos k_{y}-\cos k_{x}\cos2k_{y}\right)\,,\\
\varepsilon_{45} & = & 4t_{xy}^{45}\sin k_{x}\sin k_{y}+4t_{xxyy}^{45}\sin2k_{x}\sin2k_{y}\,.\label{eq:M45}
\end{eqnarray}

Here $\epsilon_{i}$ are the onsite energies associated with each
orbital and $t_{ij}^{\mu\nu}$ are hopping parameters from orbital
$\mu$ on site $i$ to orbital $\nu$ on site $j$. The above expressions
are accompanied by constraints on the coefficients $t_{ij}^{\mu\nu}$
due to tetragonal symmetry: 
\begin{eqnarray}
\begin{matrix}t_{x}^{11}=t_{y}^{22} &  & t_{y}^{11}=t_{x}^{22} &  & t_{xy}^{11}=t_{xy}^{22} &  & \quad t_{xx/yy}^{11}=t^{22,xx/yy}\\
t_{xxy}^{11}=t_{xyy}^{22} &  & t_{xyy}^{11}=t_{xxy}^{22} &  & t_{xxyy}^{11}=t_{xxyy}^{22} &  & t_{y}^{13}=t_{x}^{23}\\
t_{xy}^{13}=t_{xy}^{23} &  & t_{xxy/xyy}^{13}=t_{xxy/xyy}^{23} &  & t_{x}^{14}=t_{y}^{24} &  & t_{xy}^{14}=t_{xy}^{24}\\
t_{xxy}^{14}=t_{xyy}^{24} &  & t_{y}^{15}=t_{x}^{25} &  & t_{xy}^{15}=t_{xy}^{25} &  & \ t_{xxyy}^{15}=t_{xxyy}^{25}\,.
\end{matrix}
\end{eqnarray}

We are now in a position to expand the elements of $\varepsilon_{\mu\nu}(\mbf{k})$
around the $\Gamma$, $X$ and $Y$ points. At the $\Gamma$ point
the orbitals $xz$ and $yz$ dominate, corresponding to the elements
$\varepsilon_{11}$, $\varepsilon_{12}$, $\varepsilon_{21}$ and
$\varepsilon_{22}$. Similarly, at the $X$ ($Y$) point the dominant
orbitals are $yz$ ($xz$) and $xy$. To obtain these parts we expand
$\varepsilon_{22}$ ($\varepsilon_{11}$), $\varepsilon_{24}$ ($\varepsilon_{14}$)
and $\varepsilon_{44}$ around $(k_{x}+\pi,k_{y})$ ($(k_{x},k_{y}+\pi)$):
\begin{eqnarray}
h_{\Gamma} & = & \begin{pmatrix}C_{1}+C_{2}\left(k_{x}^{2}+k_{y}^{2}\right)+C_{3}\left(k_{x}^{2}-k_{y}^{2}\right) & C_{4}k_{x}k_{y}\\
C_{4}k_{x}k_{y} & C_{1}+C_{2}\left(k_{x}^{2}+k_{y}^{2}\right)-C_{3}\left(k_{x}^{2}-k_{y}^{2}\right)
\end{pmatrix}\\
h_{X} & = & \begin{pmatrix}C_{5}+C_{6}\left(k_{x}^{2}+k_{y}^{2}\right)+C_{7}\left(k_{x}^{2}-k_{y}^{2}\right) & -iv_{X}(\mbf{k})\\
iv_{X}(\mbf{k}) & C_{11}+C_{12}\left(k_{x}^{2}+k_{y}^{2}\right)+C_{13}\left(k_{x}^{2}-k_{y}^{2}\right)
\end{pmatrix}\\
h_{Y} & = & \begin{pmatrix}C_{5}+C_{6}\left(k_{x}^{2}+k_{y}^{2}\right)-C_{7}\left(k_{x}^{2}-k_{y}^{2}\right) & -iv_{Y}(\mbf{k})\\
iv_{Y}(\mbf{k}) & C_{11}+C_{12}\left(k_{x}^{2}+k_{y}^{2}\right)-C_{13}\left(k_{x}^{2}-k_{y}^{2}\right)
\end{pmatrix}\,,
\end{eqnarray}
where 
\begin{eqnarray}
v_{X}(\mbf{k}) & = & C_{8}k_{y}+C_{9}k_{y}\left(k_{y}^{2}+3k_{x}^{2}\right)-C_{10}k_{y}\left(k_{x}^{2}-k_{y}^{2}\right)\\
v_{Y}(\mbf{k}) & = & -C_{8}k_{x}-C_{9}k_{x}\left(k_{x}^{2}+3k_{y}^{2}\right)-C_{10}k_{x}\left(k_{x}^{2}-k_{y}^{2}\right)
\end{eqnarray}
As a function of the tight-binding parameters, the constants $C_{1},\ldots,C_{13}$
are 
\begin{eqnarray}
C_{1} & = & \epsilon_{\Gamma}=\epsilon_{xz/yz}+2\left(t_{x}^{11}+t_{y}^{11}\right)+4\left(t_{xxy}^{11}+t_{xyy}^{11}+t_{xy}^{11}+t_{xxyy}^{11}\right)\\
C_{2} & = & 2\frac{1}{2m_{\Gamma}}=-\frac{1}{2}\left(t_{x}^{11}+t_{y}^{11}\right)-5\left(t_{xxy}^{11}+t_{xyy}^{11}\right)-2t_{xy}^{11}-8t_{xxyy}^{11}\\
C_{3} & = & b=\frac{1}{2}\left(t_{y}^{11}-t_{x}^{11}\right)+3\left(t_{xyy}^{11}-t_{xxy}^{11}\right)-4t_{xx/yy}^{11}\\
C_{4} & = & 4c=-4\left(t_{xy}^{12}+4t_{xxyy}^{12}+4t_{xxy/xyy}^{12}\right)\\
C_{5} & = & \epsilon_{1}=\epsilon_{xz/yz}+2\left(t_{x}^{11}-t_{y}^{11}\right)-4\left(t_{xxy}^{11}-t_{xyy}^{11}+4t_{xy}^{11}-4t_{xxyy}^{11}\right)\\
C_{6} & = & 2\frac{1}{2m_{1}}=\frac{1}{2}\left(t_{y}^{11}-t_{x}^{11}\right)+5\left(t_{xxy}^{11}-t_{xyy}^{11}\right)+2t_{xy}^{11}-8t_{xxyy}^{11}\\
C_{7} & = & a_{1}=\frac{1}{2}\left(t_{x}^{11}+t_{y}^{11}\right)-3\left(t_{xxy}^{11}+t_{xyy}^{11}\right)+4t_{xx/yy}^{11}\\
C_{8} & = & 2v=2\left(t_{y}^{14}+2t_{xy}^{14}-4t_{xxy}^{14}\right)\\
C_{9} & = & 2p_{1}=-\frac{1}{12}t_{y}^{24}-\frac{25}{6}t_{xy}^{24}+\frac{7}{3}t_{xyy}^{24}\\
C_{10} & = & 2p_{2}=-\frac{1}{4}t_{y}^{24}+3t_{xyy}^{24}\\
C_{11} & = & \epsilon_{3}=\epsilon_{xy}+4\left(-t_{xy}^{44}+t_{xx/yy}^{44}+t_{xxyy}^{44}\right)\\
C_{12} & = & 2\frac{1}{2m_{3}}=2\left(t_{xy}^{44}-2t_{xx/yy}^{44}-4t_{xxyy}^{44}\right)\\
C_{13} & = & a_{3}=t_{x/y}^{44}-6t_{xxy/xyy}^{44}\,.
\end{eqnarray}
The overall minus sign in the coefficient $C_{4}$ arises due to the
minus sign in the definition of the spinor in Eq. (\ref{psi_Gamma}).
The coefficients can be obtained either by using the relations above
with the coefficients $t_{ij}^{\mu\nu}$ determined from tight-binding
fits to DFT calculations, or by directly fitting the coefficients
to DFT calculations. 
\end{widetext}

\section{Spin-orbit coupling in orbital basis}

\label{app:spin_orbit}

Here we express the standard spin-orbit coupling term $\lambda\mbf{S}\cdot\mbf{L}$
in the orbital basis, which leads to Eqs. (\ref{Gamma_SOC})-(\ref{M2_SOC})
of the main text. Denote the eigenstates of $L_{z}$ by $|m\rangle$
where $m=-L,\ldots,L$. The spin-orbit Hamiltonian is then 
\begin{eqnarray}
H_{\text{SOC}}=\sum_{\substack{mn\\
\alpha\beta
}
}\langle m\alpha|\lambda\mbf{S}\cdot\mbf{L}|n\beta\rangle d_{m\alpha}^{\dagger}d_{n\beta}\,.
\end{eqnarray}
where $d_{m\alpha}^{\dagger}$ creates an electron with spin $\alpha$
and angular momentum projection $m$. Using the fact that $\mbf{S}\cdot\mbf{L}=L_{z}S_{z}+\frac{1}{2}(L^{+}S^{-}+L^{-}S^{+})$
and 
\begin{eqnarray}
S_{z}|n\alpha\rangle=\pm\frac{1}{2}|n\alpha\rangle\,,\quad S^{\pm}|n\alpha\rangle=\delta_{\alpha,\mp\frac{1}{2}}|n,\alpha\pm1\rangle\,,
\end{eqnarray}
the Hamiltonian becomes 
\begin{eqnarray}
H_{\text{SOC}} & = & \frac{\lambda}{2}\sum_{mn}\Big[\langle m|L_{z}|n\rangle d_{m\uparrow}^{\dagger}d_{n\uparrow}-\langle m|L_{z}|n\rangle d_{m\downarrow}^{\dagger}d_{n\downarrow}\nonumber \\
 & + & \langle m|L^{+}|n\rangle d_{m\downarrow}^{\dagger}d_{n\uparrow}+\langle m|L^{-}|n\rangle d_{m\uparrow}^{\dagger}d_{n\downarrow}\Big]\nonumber \\
 & = & \frac{\lambda}{2}\sum_{\substack{mn\\
\alpha\beta
}
}A_{mn}^{\alpha\beta}d_{m\alpha}^{\dagger}d_{n\beta}\,,\label{eq:app:SOC_hamilton}
\end{eqnarray}
with the matrix elements: 
\begin{eqnarray}
A_{mn}^{\uparrow\uparrow} & = & -A_{mn}^{\downarrow\downarrow}=\langle m|L_{z}|n\rangle=n\delta_{mn}\\
A_{mn}^{\downarrow\uparrow} & = & \sqrt{(L-n)(L+n+1)}\delta_{m,n+1}\\
A_{mn}^{\uparrow\downarrow} & = & \sqrt{(L+n)(L-n+1)}\delta_{m,n-1}\,.
\end{eqnarray}
We can transform the Hamiltonian in Eq. \ref{eq:app:SOC_hamilton}
to the basis spanned by the cubic harmonics (i.e. the orbital basis)
using 
\begin{eqnarray}
d_{m\alpha} & = & \sum_{\mu}U_{m\mu}c_{\mu\alpha}\label{eq:trans_d}\\
d_{m\alpha}^{\dagger} & = & \sum_{\mu}U_{m\mu}^{\ast}c_{\mu\alpha}^{\dagger}\,,\label{eq:trans_d_dagger}
\end{eqnarray}
where $U_{m\mu}\equiv\langle m|\mu\rangle$ and $|\mu\rangle$ is
the basis states in the space of cubic harmonics. The transformed
Hamiltonian is 
\begin{eqnarray}
H_{\text{SOC}}=\frac{\lambda}{2}\sum_{\substack{\mu\nu\\
\alpha\beta
}
}\widetilde{A}_{\mu\nu}^{\alpha\beta}c_{\mu\alpha}^{\dagger}c_{\nu\beta}\,,\label{eq:app:orbita_basis_H}
\end{eqnarray}
with $\widetilde{A}^{\alpha\beta}=U^{\dagger}A^{\alpha\beta}U$.

To proceed we specialize to the case $L=2$, resulting in the well-known
$d$-orbitals. The basis states are $\langle\mu|=\langle xz|,\langle yz|,\langle xy|,\langle x^{2}-y^{2}|,\langle z^{2}|$
with 
\begin{eqnarray}
 &  & \langle xz|=\frac{1}{\sqrt{2}}\left(-\langle1|+\langle-1|\right)\\
 &  & \langle yz|=\frac{i}{\sqrt{2}}\left(-\langle1|-\langle-1|\right)\\
 &  & \langle xy|=\frac{i}{\sqrt{2}}\left(-\langle-2|+\langle2|\right)\\
 &  & \langle x^{2}-y^{2}|=\frac{1}{\sqrt{2}}\left(\langle-2|+\langle2|\right)\\
 &  & \langle z^{2}|=\langle0|\,,
\end{eqnarray}
and $U^{\dagger}$ can be read off as the coefficients of these equations.
Thus, we find $U$ to be 
\begin{eqnarray}
U=\frac{1}{\sqrt{2}}\begin{pmatrix}0 & 0 & -i & 1 & 0\\
-1 & i & 0 & 0 & 0\\
0 & 0 & 0 & 0 & \sqrt{2}\\
1 & i & 0 & 0 & 0\\
0 & 0 & i & 1 & 0
\end{pmatrix}\,,
\end{eqnarray}
yielding the three independent $\widetilde{A}$-matrices: 
\begin{eqnarray}
\widetilde{A}^{\uparrow\uparrow} & = & \begin{pmatrix}0 & -i & 0 & 0 & 0\\
i & 0 & 0 & 0 & 0\\
0 & 0 & 0 & i2 & 0\\
0 & 0 & -i2 & 0 & 0\\
0 & 0 & 0 & 0 & 0
\end{pmatrix}\,,\\
\widetilde{A}^{\downarrow\uparrow} & = & \begin{pmatrix}0 & 0 & i & 1 & -\sqrt{3}\\
0 & 0 & 1 & -i & -i\sqrt{3}\\
-i & -1 & 0 & 0 & 0\\
-1 & i & 0 & 0 & 0\\
\sqrt{3} & i\sqrt{3} & 0 & 0 & 0
\end{pmatrix}\,,\\
\widetilde{A}^{\uparrow\downarrow} & = & \begin{pmatrix}0 & 0 & i & -1 & \sqrt{3}\\
0 & 0 & -1 & -i & -i\sqrt{3}\\
-i & 1 & 0 & 0 & 0\\
1 & i & 0 & 0 & 0\\
-\sqrt{3} & i\sqrt{3} & 0 & 0 & 0
\end{pmatrix}\,.\\
\end{eqnarray}

Considering only the $|xz\rangle$, $|yz\rangle$ and $|xy\rangle$
orbitals, corresponding to the upper left $3\times3$ blocks in the
above matrices results in the restricted matrix $\widetilde{A'}_{\mu'\nu'}^{\alpha\beta}$:
\begin{eqnarray}
\widetilde{A'}=\begin{pmatrix}0 & -i & 0 & 0 & 0 & i\\
i & 0 & 0 & 0 & 0 & -1\\
0 & 0 & 0 & -i & 1 & 0\\
0 & 0 & i & 0 & i & 0\\
0 & 0 & 1 & -i & 0 & 0\\
-i & -1 & 0 & 0 & 0 & 0
\end{pmatrix}\,,
\end{eqnarray}

Decomposing it into spin and orbital parts gives: 
\begin{eqnarray}
\widetilde{A'} & = & \begin{pmatrix}0 & 0 & i\\
0 & 0 & 0\\
-i & 0 & 0
\end{pmatrix}\otimes\sigma^{x}+\begin{pmatrix}0 & 0 & 0\\
0 & 0 & -i\\
0 & i & 0
\end{pmatrix}\otimes\sigma^{y}\nonumber \\
 & + & \begin{pmatrix}0 & -i & 0\\
i & 0 & 0\\
0 & 0 & 0
\end{pmatrix}\otimes\sigma^{z}\,.
\end{eqnarray}

Finally, applying this expression to the Hamiltonian (\ref{eq:app:orbita_basis_H})
gives \cite{Sigrist00,cvetkovic01}: 
\begin{eqnarray}
\frac{\lambda}{2}\sum_{\mu'\nu'}\widetilde{A'}_{\mu'\nu'}c_{\mu'}^{\dagger}c_{\nu'} & = & i\frac{\lambda}{2}c_{xz,\alpha}^{\dagger}\sigma_{\alpha\beta}^{x}c_{xy,\beta}+\text{h.c.}\nonumber \\
 &  & +i\frac{\lambda}{2}c_{xy,\alpha}^{\dagger}\sigma_{\alpha\beta}^{y}c_{yz,\beta}+\text{h.c.}\nonumber \\
 &  & +i\frac{\lambda}{2}c_{yz,\alpha}^{\dagger}\sigma_{\alpha\beta}^{z}c_{xz,\beta}+\text{h.c.}\,,
\end{eqnarray}
which leads to Eqs. (\ref{Gamma_SOC})--(\ref{M2_SOC}).

\section{Evaluation of Two-loop Diagrams}

\label{app:two_loop_diagrams}

\begin{figure}[b]
\centering \includegraphics[width=0.45\textwidth]{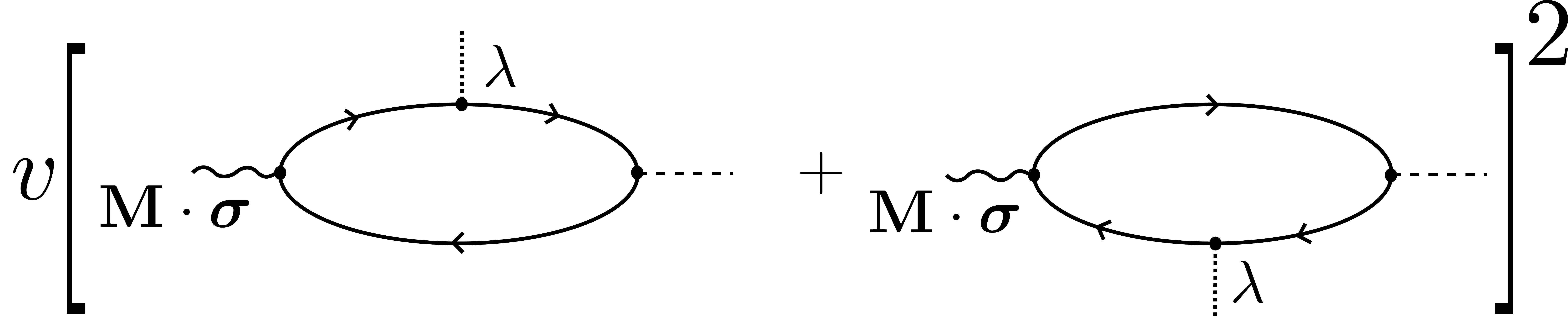}
\protect\protect\protect\protect\caption{\label{fig:irreducible_diagrams} Illustration of the decomposition
of a two-loop diagram into irreducible diagrams.}
\end{figure}

Diagrams of the two-loop type, as shown in Fig. \ref{fig:diagram_examples}a,
can be split into irreducible diagrams. As a result, all two-loop
diagrams for a given interaction can be obtained by squaring the sum
of irreducible diagrams, as illustrated in Fig. \ref{fig:irreducible_diagrams}.

To illustrate the cancelation of the two diagrams in the sum, we choose
the interaction $v_{17}$ and the order parameter $\mbf{M}_{1}$.
In this case, the sum in the brackets in Fig. \ref{fig:irreducible_diagrams}
is, for the SOC leg related to the $\Gamma$ spinor: 
\begin{eqnarray}
 &  & \frac{\lambda}{2}\sum_{\mbf{k},n}M_{1,i}\text{tr}[\sigma^{z}\sigma^{i}]\text{tr}[\tau^{+}G_{\Gamma}\tau^{y}G_{\Gamma}\tilde{\tau}^{1}G_{X}]\nonumber \\
 &  & \qquad+\frac{\lambda}{2}\sum_{\mbf{k},n}M_{1,i}\text{tr}[\sigma^{z}\sigma^{i}]\text{tr}[\tau^{-}G_{X}\tilde{\tau}^{1}G_{\Gamma}\tau^{y}G_{\Gamma}]\nonumber \\
= &  & \lambda\sum_{\mbf{k},n}M_{1,z}\Big([G_{\Gamma}\tau^{y}G_{\Gamma}]_{21}[G_{X}]_{11}\nonumber \\
 &  & \qquad+[G_{\Gamma}\tau^{y}G_{\Gamma}]_{12}[G_{X}]_{11}\Big)\,,
\end{eqnarray}

This term vanishes since $G_{\Gamma}\tau^{y}G_{\Gamma}$ is an antisymmetric
matrix. For the contribution from the diagrams with the SOC leg related
to the $X/Y$ spinors, we find: 
\begin{eqnarray}
-i &  & \lambda\sum_{\mbf{k},n}\Big(M_{1,x}\text{tr}\left(\tilde{\tau}^{1}G_{Y}\tau^{-}G_{X}\tilde{\tau}^{1}G_{\Gamma}\right)\nonumber \\
 &  & \qquad+M_{1,y}\text{tr}\left(\tilde{\tau}^{1}G_{Y}\tau^{+}G_{X}\tilde{\tau}^{1}G_{\Gamma}\right)\Big)\nonumber \\
=-i &  & \lambda\sum_{\mbf{k},n}\Big(M_{1,x}[G_{Y}]_{12}[G_{X}]_{11}[G_{\Gamma}]_{11}\nonumber \\
 &  & \qquad+M_{1,y}[G_{Y}]_{11}[G_{X}]_{21}[G_{\Gamma}]_{11}\Big)\,,
\end{eqnarray}
which is also zero as the off-diagonal elements of the Green functions
$G_{X}$ and $G_{Y}$ are odd functions of $\mbf{k}$. Similar arguments
apply in the case when the electron-electron interaction is given
by either $v_{13}$, $v_{15}$ or $v_{19}$ and when the magnetic
order parameter is $\mbf{M}_{2}$. Thus, all contributions from the
two-loop diagrams vanish, and we are left with only the one-loop diagrams
shown in Fig. \ref{fig:diagram_examples}b.

\section{Diagrams contributing to $w$}

\label{app:w_coeff}

\begin{figure}
\centering \includegraphics[width=0.25\textwidth]{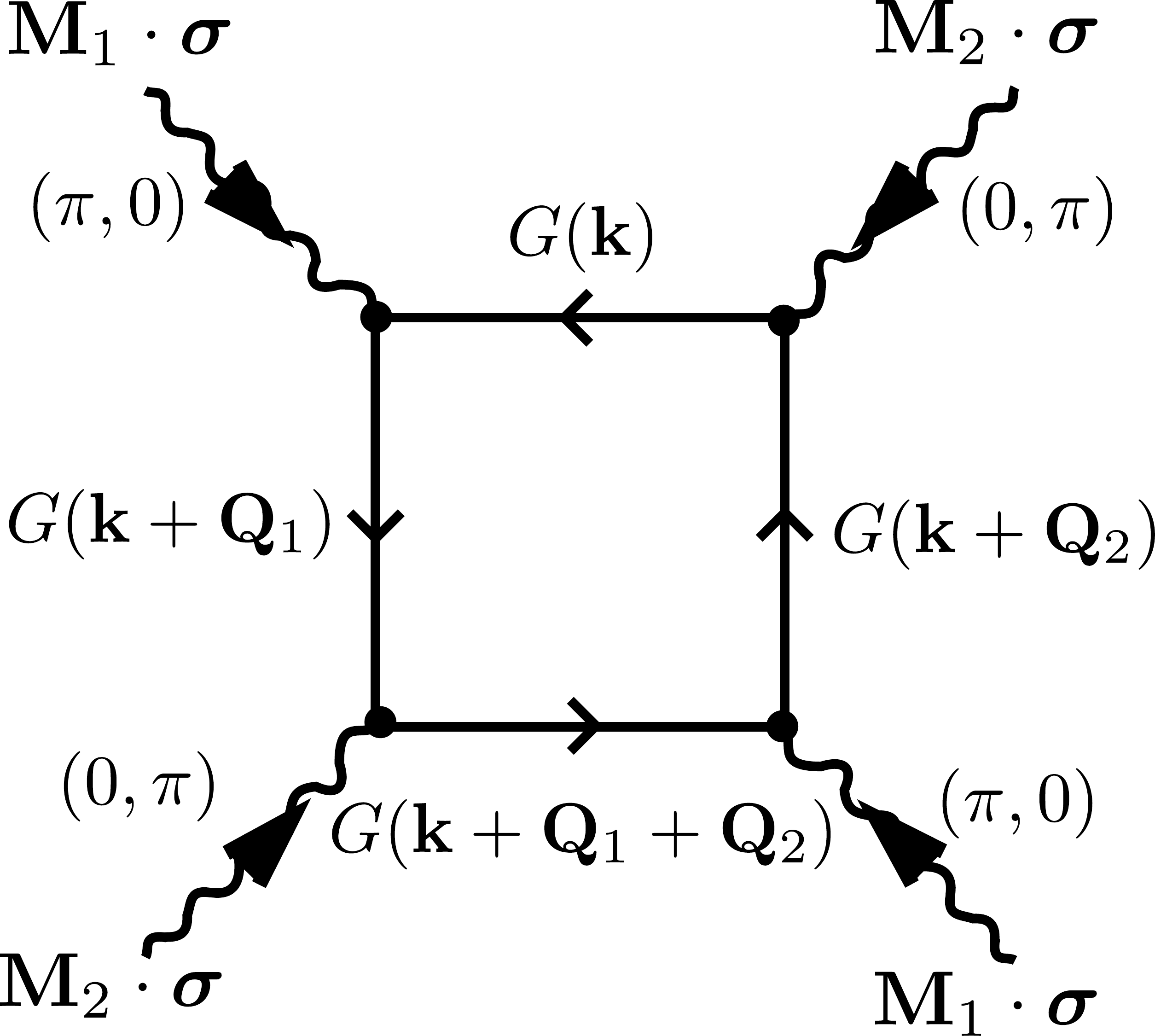}
\protect\protect\caption{\label{fig:w_diagram} Illustration of a generic diagram contributing
to the coefficient $w$. Here $\mbf{Q}_{1}=(\pi,0)$ and $\mbf{Q}_{2}=(0,\pi)$.
On the right diagram we have imposed momentum conservation at each
vertex, resulting in the appearance of $G(\mbf{k}+\mbf{Q}_{1}+\mbf{Q}_{2})$.}
\end{figure}

In this appendix we explain in more details the statement made in
the main text concerning the vanishing of the quartic coefficient
$w$ in Eq. (\ref{eqF_4_aux}). Let us consider a generic diagram
which would contribute to the coefficient $w$, as shown in Fig. \ref{fig:w_diagram}.

Note that we do not specify any vertices, i.e. the coupling between
the electrons and the SDW order parameters do not necessarily arise
from Eq. \ref{H_SDW}. Due to the Pauli matrix contraction, the diagram
must have alternating $\mathbf{M}_{1}$ and $\mathbf{M}_{2}$ legs
in order for it to contribute to $w$. Indeed, performing the trace
over spin indices gives: 
\begin{eqnarray}
M_{1}^{i}M_{2}^{j}M_{1}^{k}M_{2}^{l}\ \mathrm{tr}\left(\sigma^{i}\sigma^{j}\sigma^{k}\sigma^{l}\right)\,.
\end{eqnarray}

Using Eq. \ref{eq_Pauli}, we find: 
\begin{eqnarray}
2\left(\mbf{M}_{1}\cdot\mbf{M}_{2}\right)^{2}-\mbf{M}_{1}^{2}\mbf{M}_{2}^{2}\,,
\end{eqnarray}
thus resulting in a $\left(\mbf{M}_{1}\cdot\mbf{M}_{2}\right)^{2}$
term. This contrasts to the third diagram in Fig. \ref{fig:fourth_order_diagrams},
which gives no contribution of the form $\left(\mbf{M}_{1}\cdot\mbf{M}_{2}\right)^{2}$
after tracing over the Pauli matrices.

Let us now consider the internal lines of the diagram. Since $\mbf{M}_{1}$
carries momentum $(\pi,0)$ and $\mbf{M}_{2}$ carries momentum $(0,\pi)$,
the only way for momentum to be conserved is if one of the lines corresponds
to a propagator with momentum $\mathbf{Q}_{1}+\mathbf{Q}_{2}=(\pi,\pi)$.
However, in the absence of a Fermi pocket at $M=\left(\pi,\pi\right)$
(of the unfolded Brillouin zone), this will be an off-shell contribution.
Thus, contributions to $w$ must arise from the electronic states
near $M=(\pi,\pi)$. 
\end{document}